\documentclass[11pt]{iopart}

\usepackage{iopams}
\usepackage{setstack}
\usepackage{amscd}
\usepackage{amsfonts}
\usepackage{amssymb}
\usepackage{graphicx}
\usepackage{hyperref}

\def\be{\begin{equation}}
\def\ee{\end{equation}}
\def\ba{\begin{eqnarray}}
\def\ea{\end{eqnarray}}
\def\bc{\begin{center}}
\def\ec{\end{center}}
\newcommand{\narx}{\textit{Preprint} }
\newcommand{\arx}[1]{(\textit{Preprint} #1)}

\begin{document}

\title{Dark energy and cosmological solutions in
second-order string gravity}

\author{Gianluca Calcagni}
\address{Department of Physics, Gunma National College of 
Technology, Gunma 371-8530, Japan}
\ead{calcagni@nat.gunma-ct.ac.jp}

\author{Shinji Tsujikawa}
\address{Department of Physics, Gunma National College of 
Technology, Gunma 371-8530, Japan}
\ead{shinji@nat.gunma-ct.ac.jp}

\author{M Sami\footnote{On leave from: Department of Physics, Jamia Millia, New Delhi-110025.}}
\address{IUCAA, Post Bag 4, Ganeshkhind,
Pune 411 007, India}
\ead{sami@iucaa.ernet.in}

\begin{abstract}
We study the cosmological evolution based upon a $D$-dimensional action
in low-energy effective string theory in the presence of second-order
curvature corrections and a modulus scalar field (dilaton or compactification modulus).
A barotropic perfect fluid coupled to the scalar field is also allowed.
Phase space analysis and the stability of asymptotic 
solutions are performed for a number of models which include 
($i$) fixed scalar field, ($ii$) linear dilaton in string 
frame, and ($iii$) logarithmic modulus in Einstein frame. 
We confront analytical solutions with observational constraints for deceleration parameter 
and show that Gauss-Bonnet gravity (with no matter fields) may not 
explain the current acceleration of the universe.
We also study the future evolution of the universe using the GB 
parametrization and find that big rip singularities can be avoided 
even in the presence of a phantom fluid because of the balance between 
the fluid and curvature corrections. 
A non-minimal coupling between the fluid and the modulus field also opens
up the interesting possibility to avoid big rip  
regardless of the details of the fluid equation of state.
\end{abstract}


\maketitle


\section{Introduction}

During the past few years precision cosmology experiments have confirmed the big bang/inflationary 
standard scenario of a flat universe characterized by nearly scale-invariant, Gaussian
primordial density perturbations. Two thirds of the total energy density, which is close to the critical one,
is contributed by an exotic form of matter popularly known as dark energy 
(see \cite{de0}--\cite{de4} and references therein for an overview on the subject). 
Moreover, recent observations of extra-galactic supernovae treated as standardized candles 
indicate that the universe is presently undergoing a phase of accelerated expansion which
receives an independent support from CMB and large-scale structure data.
Within the framework of General Relativity, the cosmic speed-up can be fueled by an energy-momentum tensor
with negative pressure -- dark energy. The analysis of high redshift 
type Ia supernovae exhibits a marginal evidence of a possible super-acceleration \cite{rie04}.  
The latter eventuality has triggered many speculations 
on the underlying nature of dark energy, among which the phantom hypothesis 
has been under active investigations (see 
Refs.~\cite{Caldwell}--\cite{NOph} for early works on phantom).
By definition, a phantom is a fluid with pressure and energy density satisfying an equation of state 
$p <-\rho$; if the fluid is represented by a scalar field, 
its kinetic energy has a negative sign (hence phantoms and ghosts are relatives). 
This kind of contribution leads 
to a finite-time future singularity dubbed big rip. We also 
note that there exist other types of sudden future singularities even 
in the absence of the phantom \cite{Barrow}--\cite{NOT}.
An updated list of references on these subjects, including first and latest proposals, 
big rip phenomenology, and criticism, can be found in \cite{PhD}.

The search for a viable semiclassical or quantum model for the dark energy 
component and its dynamical properties is one of the topics explored
by ``string cosmology'', which attempts to understand the cosmological 
implications of string effective actions. In particular, quantum 
corrections to the Einstein--Hilbert action naturally arise for the 
closed string, bosonic or supersymmetric, and modify the gravitational interaction in a way 
which might be testable in the near future. The Gauss--Bonnet (GB) 
combination of curvature invariants with fixed dilaton is of special
relevance: ($i$) it represents the unique 5D leading-order extension of the Einstein--Hilbert action that leads to second order gravitational field equations,
and ($ii$) the highest derivative occurs in equations linearly, thereby 
ensuring a unique solution; finally ($iii$) with the GB parametrization, graviton interactions are ghost-free and spacetime perturbations 
are wave-like. However, in four dimensions, the GB term with fixed 
dilaton is topological and does not contribute to the equations of motions. 
Then, the only way to modify the cosmological evolution is to 
consider GB term in the bulk in a braneworld setup, so that the induced 4D 
Friedmann equation on the brane depends on the bulk back-reaction. 
The GB braneworld and its observational consequences, 
both in the early and recent universe, are still under 
extensive study (see \cite{PhD} for a list of references).

Second-order gravity terms can generate non-trivial effects in a four-dimensional 
spacetime even outside the braneworld lore, provided that we relax some common 
assumptions: the most important ones are fixed dilaton and GB parametrization. In the first case, it is
assumed that non-perturbative potentials arise and dilaton and other moduli freeze 
out with a minimum non-zero mass, so that to preserve the observational bounds 
for the gravitational interaction \cite{ETV,TayV}. However, a massless dilaton still can be 
compatible with observations \cite{DP1,DP2,DPV} (see also \cite{DN1,DN2}).

If the dilaton varies in spacetime, gradient terms appear and the 
higher-order curvature invariants interact with a non-constant $\alpha'$-coupling. 
Also, compactification of the 26 or 10-dimensional target space is encoded in  
residual modulus fields which in general will evolve in time. Both cases result 
in a dynamical field non-minimally coupled to gravity.

Although the Gauss--Bonnet combination of curvature invariants is ghost-free, we should 
bear in mind that in the high-energy limit the uniqueness of the GB term 
is not guaranteed and, from the point of view of string theory, 
the GB parametrization cannot be 
distinguished from the others (see references in \cite{MS}).

Dilaton and modulus cosmologies in higher-order gravity were considered, 
e.g., in \cite{MS}--\cite{PTZ}. 
As regards the recent universe, dark energy models with higher-order terms 
were inspected, e.g., in \cite{STTT} in the case of fixed moduli, 
while a dynamical dilaton as a dark energy candidate 
was studied in \cite{GPV,PT}.
Recently Nojiri \textit{et al.} \cite{NOS} investigated a dark energy scenario
 in GB gravity coupled to a dynamical scalar field
 with non-negative potential.

The understanding of the impact of higher-order curvature terms on the macrocosmic
evolution may open up the possibility to verify high-energy fundamental theories 
through cosmology. We address this issue in two ways. First, we find cosmological 
solutions in the absence of an extra fluid in the action and 
constrain them via the recent supernovae data \cite{rie04}.
It turns out that, while solutions with non-GB parametrizations can fit the 
experimental evidence, the GB solutions are in general not viable, 
unless one takes the perfect fluid into account or, in the case of 
finite-time future singularity solutions, one fine-tunes the 
parameters of the model. 
Second, we solve numerically the equations of motion with a perfect 
fluid in four dimensions and study the future behaviour of its energy density
and the Hubble parameter. In particular we are interested in the case 
of a phantom fluid which leads to the growth of the energy density of
the universe. Then higher-order curvature corrections become 
inevitably important around the big rip.
In fact, we will show that both second-order curvature terms and a non-minimal 
coupling between the fluid and the modulus field deeply affect the cosmological 
evolution. We shall classify the situations in which big rip singularities 
are reached or avoided.
We keep arbitrary coefficients in the second-order gravity action, 
considering then the usual Gauss--Bonnet parametrization as a special case. 
Our results are in agreement with past 
investigations in literature where comparison is possible. Also, 
this work extends previous analyses to a spacetime of arbitrary 
dimension (including the 4D case) with a (coupled) barotropic fluid being taken into account.

The paper is organized as follows. In section \ref{setup} we present a general 
action which involves curvature, scalar field, and barotropic fluid terms, 
and derive basic equations for this system.

In section \ref{dS} we take up the issue of de Sitter and inflationary solutions in the 
case of a constant scalar field. We then work out the phase space and stability analysis 
for a linearly changing dilaton in the string frame in section \ref{linear}. 
Some technical material is confined to the Appendix.

Section \ref{modu} is devoted to the logarithmic modulus case with stabilized dilaton, 
which solves either exactly or asymptotically the equations of motion.
The stability of these solutions is studied, 
together with their application as concrete models for the late time
evolution of the universe.

The properties of second-order gravity in the context of (phantom) dark energy 
are discussed in section \ref{dark}.
We show that it is possible to avoid big rip singularity in the 
presence of the Gauss--Bonnet corrections with a dynamically changing 
modulus field. The coupling $Q$ between the modulus and the fluid can also 
play important roles in determining the future fate of the universe. 
Conclusions are given in section \ref{concl}.


\section{General setup}\label{setup}

In our notation, $g_{\mu\nu}$ is the $D$-dimensional metric with signature $(-,+,\cdots,+)$ and Greek 
indices run from 0 to $d\equiv D-1$. The Riemann and Ricci tensors are 
defined as $R^\alpha_{\beta\mu\nu}\equiv\partial_\mu\Gamma^\alpha_{\beta\nu}-\partial_\nu\Gamma
^\alpha_{\beta\mu}+\Gamma^\alpha_{\mu\sigma}\Gamma^\sigma_{\beta\nu}-\Gamma^\alpha_{\nu\sigma}\Gamma^\sigma_{\beta\mu}$ 
and $R_{\mu\nu}\equiv R^\alpha_{\mu\alpha\nu}$, respectively, where 
$\Gamma^\alpha_{\beta\nu}$ is the Christoffel symbol. 
The gravitational coupling $\kappa_D$ and the Regge slope $\alpha'$ are 
set to unity.


\subsection{General action}

The gravity+matter action we begin from is (e.g., \cite{HN,CHC})
\be\label{act}
\fl {\cal S}=\int \rmd^Dx \sqrt{-g}\left[\frac12 
f(\phi,R)-\frac12\omega(\phi)(\nabla\phi)^2-V(\phi)+\xi(\phi){\cal 
L}_c^{(\phi)}+{\cal L}_\rho^{(\phi)}\right],
\ee
where $g$ is the determinant of the $D$-dimensional metric, $\phi$ is 
a scalar field corresponding either to the dilaton or to 
another modulus, and $f$ is a generic function of the 
scalar field and the Ricci scalar $R$.
$\omega$, $\xi$ and $V$ are functions of 
$\phi$. In this paper we do not consider the cosmological dynamics 
in the presence of the field potential $V$, but we include this term 
in deriving basic equations.
${\cal L}_\rho^{(\phi)}$ is the Lagrangian of a $D$-dimensional
perfect fluid with energy density $\rho$ and pressure $p$.
Later on we shall assume that the barotropic index $w\equiv p/\rho$ is a constant.
In general the fluid will be coupled to the scalar field $\phi$.

Finally, $\alpha'$-order quantum corrections are encoded in 
the term
\be 
\label{Lc}
{\cal L}_c^{(\phi)} 
=a_1R_{\alpha\beta\mu\nu}R^{\alpha\beta\mu\nu}
+a_2R_{\mu\nu}R^{\mu\nu}
+a_3R^2+a_4(\nabla\phi)^4,
\ee
where $a_i$ are coefficients depending on the string model one is 
considering. In section \ref{sce} we will specify the functions $f$, 
$\omega$, $V$ and $\xi$ together with the coefficients $a_i$ 
according to the scenarios of interest.

We assume that the target spacetime is described by a flat 
Friedmann-Robertson-Walker metric
\be\label{FRW}
\rmd s^2=-N^2(t)\rmd t^2+
a^2(t)\sum_{i=1}^d (\rmd x^i)^2,
\ee
where $N(t)$ is the lapse function ($N=1$ in synchronous gauge) and 
$a(t)$ is the scale factor
describing the physical size of the universe.
The size of the causal connected region centered on the 
observer is given by the inverse of the Hubble parameter 
$H\equiv\dot{a}/(Na)$. In the following, dots and primes denote 
derivatives with respect to synchronous time $t$ and $\phi$, 
respectively. 


\subsection{Equations of motion}

With the metric (\ref{FRW}), the Riemann invariants read
\ba
R_{\alpha\beta\mu\nu}R^{\alpha\beta\mu\nu} =
2d\left[2\left(H^2+\frac{\dot{H}}{N}\right)^2+(d-1)H^4\right],\\
R_{\mu\nu}R^{\mu\nu} = 
d\left[d\left(H^2+\frac{\dot{H}}{N}\right)^2+\left(dH^2
+\frac{\dot{H}}{N}\right)^2\right],\\
R = d\left[(d+1)H^2+2\frac{\dot{H}}{N}\right],
\ea
where we used
\be
\frac{\ddot{a}}{N^2a}-\frac{\dot{N}}{N}\frac{H}{N}
=H^2+\frac{\dot{H}}{N}.
\ee
For simplicity we shall limit the discussion to a
homogeneous scalar field $\phi(t)$. 
Then the spatial volume can be integrated out from 
the measure in equation (\ref{act}), which we rewrite as
\be
{\cal S}=\int \rmd t Na^d  
\left[{\cal L}_g+{\cal L}_\rho^{(\phi)}\right].
\ee
The Hamiltonian constraint 
$\delta {\cal S}/\delta N|_{N=1}=0$ is
\be
\left.\left({\cal L}_g+N\frac{\partial {\cal L}_g}{\partial 
N}-dHN\frac{\partial {\cal L}_g}{\partial 
\dot{N}}-N\frac{d}{dt}\frac{\partial {\cal L}_g}{\partial 
\dot{N}}-\rho\right)\right|_{N=1}=0,
\ee
giving the Friedmann equation
\ba
\label{FReq1}
d(d-1)FH^2 = RF-f-2dH\dot{F}+2(\rho+\rho_\phi+\xi\rho_c),
\ea
where
\ba
\label{FReq2}
\fl \rho_\phi &=& \frac12\omega\dot{\phi}^2+V, \\
\label{FReq3}
\fl\rho_c &=& 3a_4\dot{\phi}^4-d[4c_1\Xi H^3+(d-3)c_1H^4+c_2(2\Xi 
H\dot{H}+2dH^2\dot{H}-\dot{H}^2+2H\ddot{H})].
\ea
Here  $F\equiv \partial f/\partial R$, $\Xi\equiv \dot{\xi}/\xi$, and
\begin{numparts}\ba
c_1  &\equiv& 2a_1+da_2+d(d+1)a_3,\label{ci1}\\
c_2  &\equiv& 4a_1+(d+1)a_2+4da_3.\label{ci2}
\ea\end{numparts}
Note that all three energy densities, $\rho$, $\rho_{\phi}$ and 
$\rho_{c}$, depend on the field $\phi$. In four dimensions ($d=3$), the 
coefficients (\ref{ci1})--(\ref{ci2}) read $c_1=2a_1+3a_2+12a_3$ and 
$c_2=4(a_1+a_2+3a_3)$, while the $H^4$ term in $\rho_c$ vanishes. At 
low energy it was shown that the unique higher-order gravitational 
Lagrangian giving a theory without ghosts is the Gauss--Bonnet 
one ($a_1=a_3=1$, $a_2=-4$, $a_4=-1$). In this case, $c_2$ vanishes identically 
while $c_1=2+d(d-3)$. With fixed dilaton coupling ($\Xi=0$)
equation (\ref{FReq1}) reduces to the standard Friedmann 
equation in four dimensions, 
in agreement with the fact that the GB term is topological
when $d=3$. In three dimensions ($d=2$), the GB higher-derivative 
contribution vanishes identically except for the $\dot{\phi}^4$ term.

The continuity equation for the dark energy fluid contains a source 
term given by the coupling between this fluid and the string scalar 
field. We choose the covariant coupling considered in 
\cite{wet95,ame99,ame00}: 
$\delta {\cal S}_\rho/\delta\phi
=-\sqrt{-g}\,Q(\phi)\rho$, where ${\cal S}_\rho=\int \rmd^Dx 
\sqrt{-g} {\cal L}_\rho^{(\phi)}$ and $Q(\phi)$ is an unknown 
function which we shall set to a constant later. 
In synchronous gauge we have
\be\label{drho}
\dot{\rho}+dH\rho(1+w)=Q\dot{\phi}\rho,
\ee
while the equation of motion for the field $\phi$ is
\be\label{pheom}
\fl
\omega(\ddot{\phi}+dH\dot{\phi})+V'-\xi'{\cal 
L}_c^{(\phi)}+4a_4\xi\dot{\phi}^2(3\ddot{\phi}
+dH\dot{\phi}+\Xi\dot{\phi})
+\left(\dot{\omega}\dot{\phi}-\omega'\frac{\dot{\phi}^2}{2}
-\frac{f'}{2}\right)=-Q\rho,
\ee
where the Lagrangian of the quantum correction 
is written as
\be
{\cal L}_c^{(\phi)} = 
d\left[(d+1)c_1H^4+4c_1H^2\dot{H}+c_2\dot{H}^2\right]
+a_4\dot{\phi}^4.
\ee
Equations (\ref{FReq1})--(\ref{pheom}) are the master 
equations of the physical system under study.


\subsection{Dilaton and modulus scenarios}\label{sce}

In order to reproduce general relativity at low energy, we impose that 
the leading term is given by the Einstein--Hilbert linear 
Lagrangian, i.e., $f(\phi,R) = F(\phi)R$.

The massless dilaton field arises in the string loop expansion of the low-energy 
effective theory, its vacuum expectation value being directly related to 
the string coupling (e.g., \cite{pol}). It is possible to consider the 
physical properties of the dilaton in two different frames (see, e.g., 
\cite{mae89,GV,GV02}). In the ``string frame'' (subscript $S$), 
the dilatonic action corresponds to
\begin{numparts}\ba
F &=& -\omega =e^{-\phi},\label{eh}\\
V &=& 0\,,\\
\xi &=& \frac{\lambda}{2} e^{-\phi},\label{tree}
\ea\end{numparts}
where $\lambda= 1/4, 1/8$ for the bosonic and heterotic 
string, respectively, whereas $\lambda=0$ in the Type II superstring. 
In fact, the above expression for $\xi$ is the tree-level term 
in the full contribution of $n$-loop corrections, given by 
\be
\xi =\frac12 \lambda \sum_{n=0} C_{n}
e^{(n-1)\phi}\,,
\ee
where $C_{0}=1$. In this work we shall take only 
the tree-level term (\ref{tree}) into account. 
Note that our function $\xi(\phi)$ is dubbed 
$-\frac12 \alpha'\lambda\xi(\phi)$ in \cite{CHC}.

One can transform to the ``Einstein frame'' 
($f=R$, subscript $E$) by making a 
conformal transformation
\be
g_{\mu\nu}^{(S)} = F^{\frac{2}{D-2}}g_{\mu\nu}^{(E)} 
=e^{\frac{2\phi}{D-2}}g_{\mu\nu}^{(E)},
\ee
so that $F_E=\omega_E=1$, while the term ${\cal L}_c^{(\phi)}$ is 
modified accordingly (note that $\xi_E=\xi_S$ in four dimensions). 
Although the analysis can be performed in any 
frame, the equivalence principle is preserved 
only in the Einstein frame (in the string frame, the dilaton coupling to gravity 
results in a varying gravitational ``constant''). One can shift from 
the string to the Einstein frame by noting that
\ba
H_E &=& 
e^{\frac{\phi}{D-2}}\left(H_S-\frac{\dot{\phi}}{D-2}\right),
\label{EiSt}\\
\dot{H}_E &=& e^{\frac{2\phi}{D-2}} 
\left[\dot{H}_S+\frac{\dot{\phi}H_S}{D-2}-\frac{\ddot{\phi}}{D-2}
-\left(\frac{\dot{\phi}}{D-2}\right)^2\right],\label{EiSt2}
\ea
where dots in the left (right) hand side denote derivatives with 
respect to the Einstein (string) time coordinate. 
Two frames coincide when the dilaton is fixed.

In the string frame, the equations of motion read:
\ba
\fl d(d-1)H^2 = (2dH-\dot{\phi})\dot{\phi}+2e^\phi\rho+\lambda\rho_c,\\
\fl -Qe^\phi\rho =\ddot{\phi}+dH\dot{\phi}+\frac{\lambda}{2}{\cal 
L}_c^{(\phi)}+2a_4\lambda\dot{\phi}^2(3\ddot{\phi}
+dH\dot{\phi}-\dot{\phi}^2)
+\frac12\left(\dot{\phi}^2+R\right),
\ea
where one has $\Xi=-\dot{\phi}$ in the definition of $\rho_c$.

In general, other moduli appear whenever a submanifold of the target 
spacetime is compactified with compactification radii described by the expectation 
values of the moduli themselves. In the case of a single modulus (one common 
characteristic length) and heterotic string ($\lambda=1/8$), 
the four-dimensional action corresponds to \cite{ART}
\begin{numparts}\ba
\label{modu1}
F &=& 1,\\
\omega &=& 3/2,\\
a_4 &=& 0,\\
\xi &=& -\frac{\delta}{16}\ln [2e^\phi\eta^4(ie^\phi)],
\label{modu2}
\ea\end{numparts}
where $\eta$ is the Dedekind function and $\delta$ is a constant proportional 
to the 4D trace anomaly and depending on the number of chiral, vector, 
and spin-$3/2$ massless supermultiplets of the $N=2$ sector of the theory. 
In general it can be either positive or negative, 
but it is positive for the theories in which not
too many vector bosons are present. 
Again the scalar field corresponds to a flat direction 
in the space of nonequivalent vacua and $V=0$.
At large $\phi$ the last equation can be approximated as
\ba
\xi &\approx& \xi_0 \cosh\phi,\label{ximodu1}\\
\xi_0 &\equiv& \frac{\pi \delta}{24},\label{ximodu2}
\ea
which we shall use instead of the exact expression. 
In fact it was shown in Ref.~\cite{YMO} that 
this approximation gives results very close to those of the exact case.

In any realistic situation both the dilaton and the modulus can appear 
in the action. However, we can consider the effect of each of them 
separately for the following reason. 
As already mentioned, a varying dilaton is strongly constrained 
by gravitational experiments.
Although there is still a possibility to allow for a non-trivial dynamical contribution, 
one can assume to freeze it out at the minimum of its non-perturbative potential. 
Once the dilaton is fixed, gravity is not coupled to the surviving modulus field. 
Also, while in the modulus case one considers the heterotic string already 
compactified to $d=3$, in the dilaton case we should start from 
the 26- or 10-dimensional action and then compactify down to four dimensions. 
To set $d=3$ into the action is equivalent to consider a stabilized modulus (or moduli) 
while keeping the dilaton as a flat direction of the theory.


\section{de Sitter and inflationary solutions: fixed scalar field}\label{dS}


\subsection{Preliminary remarks and geometrical inflation}

The search for de Sitter (dS) solutions of the higher-derivative 
equations of motion \cite{BaM} is 
important at least for two reasons. First, they can give useful 
insight as regards sensible implementations of string-motivated 
scenarios in the early-universe inflationary context \cite{BB,MO05}. 
Secondly, stable dS solutions may avoid the infall of the 
late universe in eventual finite-time singularities. In order to take 
into account the case of inflation, we shall consider small but 
non-vanishing time derivatives of the Hubble parameter. 
We introduce the slow-roll (SR) parameter 
\be\label{sr}
\epsilon \equiv -\frac{\dot{H}}{H^2},
\ee
while leaving the other SR parameters specified implicitly in the 
second-order relation $\ddot{H}=\Or(\epsilon^2)$.
Since the evolution equation for $\epsilon$ is at second-order 
in the SR parameters themselves, these are constant if small 
enough (slow-roll approximation). 
Pure dS solutions correspond to 
$\epsilon=\cdots=0$, where dots stand for other parameters of the 
SR tower. The Friedmann equation (\ref{FReq1})
can be truncated at 
lowest order in the SR parameters by neglecting $\Or(\epsilon^2)$ 
contributions.\footnote{Typical inflationary scenarios are achieved 
via the introduction of a dynamical scalar field. 
If we consider $\rho$ in the action (\ref{act}) as the energy density of the 
inflaton, then it would be trivial to find quasi de Sitter solutions. 
Here our aim is to consider the pure gravitational action with 
$\rho=0$ and see what kind of cosmological evolution 
is obtained in the presence of curvature corrections.} 
Let us study the case of a constant scalar 
field ($\dot{\phi}=0$) in order to 
present a preliminary example.

In general, (quasi) de Sitter 
solutions can exist for a non-linear $f(R)$ as in the 
Starobinsky model \cite{sta80}.
In the case of linear gravity, we can fix $F=1=\xi$ 
without loss of generality. If the 
scalar field $\phi$ vanishes or is constant with vanishing potential 
($\rho_\phi=0=\Xi$), the Friedmann equation becomes a differential 
Riccati-type equation in $H$,
\be\label{ric}
4c_2d\dot{H}+2(d-3)c_1H^2+(d-1)=0,
\ee
where we have factored out the constraint giving the trivial 
Minkowski spacetime $H=0$. 
Clearly there are no dS solutions for $d=1$ and $d=3$. 
At low energies the only invariant 
Lagrangian which is second order in the Riemann tensor and contains 
no ghosts is the Gauss--Bonnet invariant, which coincides 
with the Euler characteristic of the target manifold in four dimensions. 
As $c_2=0$, not even inflationary solutions would exist in the GB case.
Nevertheless it is worth studying the case of $c_2\neq 0$
for the reasons explained in the introduction.
Then the 4D inflationary solution is
\be\label{4Dinsol}
H(t)=-\frac{t}{6c_2},
\ee
with $c_2<0$. When $1\neq d\neq 3$, there are both dS and 
inflationary solutions. The former is given by \cite{MS}
$H^2= -(d-1)/[2(d-3)c_1]$. If $d>3$, then $c_1$ must be negative, 
which excludes the GB case. The quasi dS solution for equation 
(\ref{ric}) with $c_2\neq 0$ is
\be\label{tanh}
H(t)= \sqrt{\frac{d-1}{2(3-d)\xi_0c_1}} \tanh 
\left[\sqrt{\frac{(d-1)(3-d)c_1}{8d^2\xi_0c_2^2}}\,t\right],
\ee
which is obtained by fixing the integration constant so 
that $H(0)=0$ and by assuming $c_1,c_2<0$ 
($H(t)$ is positive and monotonic in this case).
Changing the sign of $c_1$ results in a periodic tangent behaviour.
The Hubble rate (\ref{tanh}) approaches 
a constant in the limit $t \to \infty$.
To summarize, higher-derivative terms can induce an inflationary 
phase of pure geometrical nature. 
This unusual situation\footnote{Curvature-driven accelerating models 
were already considered in, e.g., \cite{sta80}--\cite{DBD} and references therein. See also \cite{NO3} for a case which can be compatible with solar system experiments.} 
is anyway avoided in any dimension when considering a Gauss--Bonnet coupling. 
This might be another reason why to fix the coefficients $c_i$ to their GB value.

A stability analysis can be performed as follows. Let $H_c$ be a 
solution of equation (\ref{ric}) and $H=H_c+\delta H$ be another solution given
by the first plus a small time-dependent perturbation $\delta H\neq 0$. 
If the perturbation decays in time, then the 
solutions are stable. Dropping ${\cal O}(\delta H^2)$ terms in equation 
(\ref{ric}), one gets 
\be
\delta H(t) = a(t)^{(3-d)c_1/(dc_2)},
\ee
where we have used the definition of the Hubble parameter. Therefore the dS 
solution for $d>3$ exists and is stable if $c_1,c_2<0$.\footnote{This result, confirmed by the following numerical analysis, is in 
disagreement with \cite{MS}, where it was claimed that dS solutions 
are unstable regardless of the parametrization. In particular, we were not able to reproduce their equations 
(33)--(36).} Since the inflationary solutions can 
be regarded as perturbations of the dS ones, intuitively these are 
stable when dS solutions are; indeed the inflationary solution (\ref{tanh}) is stable.

In the 4D case corresponding to 
equation (\ref{4Dinsol}), the perturbation is constant and therefore 
the solution is not stable, as could be guessed by the fact that 
there is no dS background to perturb.

We caution the reader that from the point of view of quantum gravity, models with a non-GB parametrization are unstable because of the presence of ghost fields. In this respect, whenever we mention the stability issue for such models we are implicitly referring to the classical stability against linear perturbations of a classical solution within a phenomenological effective theory. The details of the latter are still far for being established, as we shall discuss in the concluding section.


\subsection{Phase space analysis}

One may wonder if the slow-roll approximation we used in the previous 
subsection is really justified. 
A priori the $\ddot{H}$ term may play a relevant role in 
the analysis, since the stability in that direction was not 
considered. In the next sections we clarify this point 
and generalize to the case of a dynamical scalar field. 
We shall consider the cases $c_2\neq 0$ and
$c_2=0$ (GB scenario) separately.

By introducing the variables 
\be
x\equiv H\,,\qquad  y\equiv\dot{H}\,,\qquad 
u\equiv\phi\,,\qquad
v\equiv\dot{\phi}\,,\qquad z\equiv\rho\,,
\ee
we obtain the phase space evolution equations
\ba
\fl\dot{x} &=& y\,,\label{eoms1} \\
\fl 
\dot{y} &=& -\frac{(d-1)Fx}{4\xi c_{2}}-\frac{\dot{F}}
{2\xi c_{2}}+\frac{2z+\omega v^2+2V}{4d\xi c_{2}x}
+\frac{3a_{4}v^4}{2dc_{2}x}\nonumber\\
\fl 
& & \qquad-\frac{4c_1\Xi x^3+(d-3)c_1x^4+c_2(2\Xi 
xy+2dx^2y-y^2)}{2c_{2}x}\,,\label{eoms2}\\
\fl 
\dot{u} &=& v\,,\label{ueq} \\
\fl 
\dot{v} &=& \frac{-d\omega xv -V'+\xi' {\cal L}_c^{(\phi)}
-4a_{4} \xi v^3 (dx+\Xi)-\dot{\omega}v+(\omega'
v^2+F'R)/2-Qz}
{\omega+12a_{4}\xi v^2}\,, \label{veq} \\
\fl 
\dot{z} &=& \left[-dx(1+w)+Qv \right]z\,,\label{zeq}
\ea
where
\ba
{\cal L}_c^{(\phi)} &=& d[(d+1)c_{1}x^4+4c_1x^2y+c_{2}y^2]+a_{4}v^4,\\
R &=& d[(d+1)x^2+2y].
\ea
In what follows we do not take into account the scalar field potential, which 
is not present in a perturbative string framework. 
In addition the coupling $Q$ is assumed to be constant.

In the GB case ($c_2=0$) differentiating the Friedmann 
equation (\ref{FReq1}) with respect to time gives the 
following equation
\ba
\fl 
2d[\dot{F}+(d-1)Fx+12c_1\dot{\xi}x^2+4(d-3)c_1\xi x^3]\dot{x} = 
(RF-f)^{\bullet}
-2d\ddot{F}x-d(d-1)\dot{F}x^2\nonumber\\
\fl 
+2\dot{z}+\dot{\omega}v^2+2\omega v\dot{v}+2vV'+
6a_4v^3(\dot{\xi}v+4\xi\dot{v})-2d(d-3)c_1\dot{\xi}x^4
-8dc_1\ddot{\xi}x^3.\label{frc20}
\ea

Let us search for a dS solution which is characterized by 
$\dot{x}=0$, $\dot{y}=0$ with a vanishing fluid ($\dot{z}=z=0$).
We find that 
\be
\label{deSidilaton}
\fl 
2\xi d(d-3) c_{1}x^4+8\xi dc_{1}\Xi x^3+d(d-1)Fx^2+
2d\dot{F}x-\omega v^2-6a_{4} \xi v^4=0\,.
\ee
When the scalar field $\phi$ is fixed with vanishing potential ($\xi=1=F$), 
equations (\ref{FReq1}) and (\ref{drho}) reduce to 
\ba
& & d(d-1)H^2=2\rho-2 
d\left[(d-3)c_{1}H^4+c_{2}(2dH^2\dot{H}
-\dot{H}^2+2H\ddot{H})\right], \\
& & \dot{\rho}+dH(1+w)\rho=0\,.
\ea
Defining the vector $\tilde{X}\equiv (x,y,z)^t$, 
the above equations of motion can be written as
\ba
\label{x}
\dot{x} &=& y\,,\\
\label{y}
\dot{y} &=& 
\frac{(3-d)c_1}{2c_2}x^3-dxy+\frac{y^2}{2x}-\frac{d-1}{4 
c_2}x+\frac{z}{2dc_2 x},\\
\label{z}
\dot{z} &=& -d(1+w)xz\,,
\ea
where we assumed that the perfect fluid has a constant barotropic index 
$w$. In the following we treat the case $w=-1$ (cosmological 
constant) separately. Also, we consider a space with $d>3$
in this section.


\subsection{Perfect fluid with $w \neq -1$}

The dS fixed point $\dot{\tilde{X}}_c=0$ for the previous set of 
equations is
\be
\label{deSitter}
x_{c}=\sqrt{\frac{(d-1)}{2(3-d)c_{1}}}\,,\qquad
y_{c}=0\,,\qquad z_{c}=0\,,
\ee
where we have considered only the expanding case.
When $d>3$ we require the condition $c_{1}<0$ for the existence of 
the dS solution. 

\begin{figure}
\bc
\includegraphics[height=3.0in,width=3.2in]{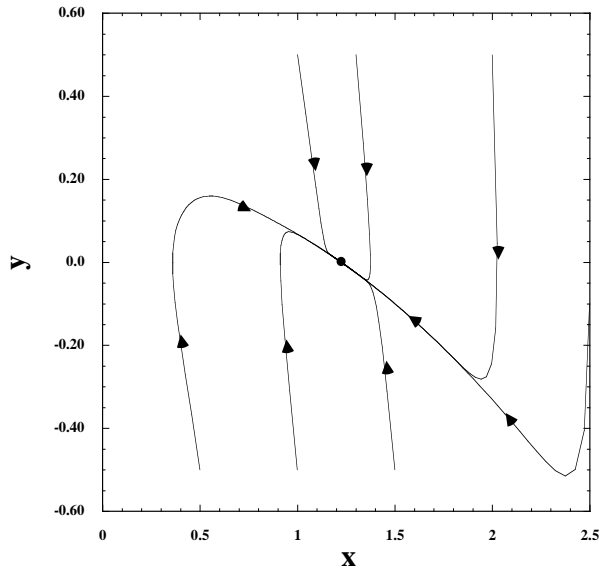}
\ec
\caption{\label{fig1}
The phase space plot for $d=4$, $\xi=1$, $F=1$, 
$w=-0.9$, $c_{1}=-1$ and
$c_{2}=-1$ when the scalar field $\phi$ is fixed.
In this case the de Sitter fixed point $(x_c, y_c, z_c)=
(1.22, 0, 0)$ is a stable attractor.
}
\end{figure}

Linearizing the equations of motion under a perturbation 
$\delta \tilde{X}=\tilde{X}-\tilde{X}_c$ 
with respect to a solution $\tilde{X}_c$, one obtains
\ba
\fl
\delta\dot{x} = \delta y,\\
\fl
\delta\dot{y} = 
\left[\frac{3(3-d)c_1}{2c_2}x_c^2-dy_c-\frac{y_c^2}{2x_c^2}
-\frac{(d-1)}{4c_2}-\frac{z_c}{2dc_2 x_c^2}\right]\delta 
x+\left(\frac{y_c}{x_c}-dx_c\right)\delta y+\frac{1}{2d c_2 
x_c}\delta z,\nonumber\\\\
\fl
\delta\dot{z} = -d(1+w)z_{c}\delta x-d(1+w)x_c\delta z.
\ea
This can be rewritten in compact notation as $\delta \dot{\tilde{X}}=\tilde{M}\delta 
\tilde{X}$, where $\tilde{M}$ is a $3 \times 3$ matrix with elements $m_{i\!j}$ and 
eigenvalues $\gamma$ defined by the Jordan constraint
\be
\label{eqL}
\fl
\gamma^3-(m_{33}+m_{22})\gamma^2+(m_{22}m_{33}-m_{21})
\gamma+(m_{21}m_{33}-m_{31}m_{23})=0\,.
\ee
For the dS fixed point given by 
equation (\ref{deSitter}) one has 
$m_{31}=-d(1+w)z_{c}=0$. In this case equation (\ref{eqL}) is factored out 
as $\gamma=m_{33}$ and $\gamma^2-m_{22}\gamma-m_{21}=0$.
Therefore we find that the eigenvalues are
\be
\gamma_{1}=-d(1+w)x_{c}\,,\qquad
\gamma_{2, 3}=
\frac12 \left[-dx_{c} \pm \sqrt{d^2x_{c}^2+
\frac{2(d-1)}{c_{2}}} \right].
\ee
This implies that the dS solution is unstable for a phantom 
fluid, since $\gamma_{1}$ is positive for $w<-1$.
When $c_{2}<0$ the real part of both $\gamma_{2}$ and 
$\gamma_{3}$ is negative. Therefore the dS expanding solution 
is stable for $w>-1$ and $c_{2}<0$, as is illustrated 
in figure \ref{fig1}.
When $c_{2}$ is 
positive we have $\gamma_1>0$ and $\gamma_2<0$, implying that the 
dS solution is a saddle point. In summary, there is a dS 
solution for $c_1<0$ and $d>3$ which is a stable attractor for 
$c_2<0$ and $w>-1$. Note that parameters with this sign choice
do not avoid either ghost-free graviton scattering amplitudes 
or naked singularity structures \cite{BD}. 

\subsection{Cosmological constant $(w = -1)$}\label{w-1}

In this subsection we shall discuss the situation in which a 
cosmological constant $\Lambda$ is present instead of a barotropic 
fluid. This corresponds to dropping equation~(\ref{z}) and  
replacing $\rho$ for $\Lambda$.
Then we find that the dS solution satisfies 
\be
\label{Lam}
\Lambda=d(d-3)c_1x_{c}^4+\frac
{d(d-1)x_{c}^2}{2}\,,\qquad y_{c}=0\,.
\ee
There exists one solution for this equation when $(d-3)c_1$ is 
positive, whereas two solutions exist for $(d-3)c_1<0$
as long as the cosmological constant satisfies $0<\Lambda< 
d(d-1)^2/[16(3-d)c_{1}]$.
In the latter case each solution $x=x_{c}$ belongs to the range 
\ba
\label{two}
0<x_{c}<x_{M} \equiv \sqrt{\frac{(d-1)}{4(3-d)c_1}}\,,
~~~{\rm and}~~~
x_{M}<x_{c}<
\sqrt{\frac{(d-1)}{2(3-d)c_1}}\,.
\ea
One can compute the eigenvalues of the $2 \times 2$ matrix for 
perturbations $\delta x$ and $\delta y$ about the dS fixed 
points given by equation~(\ref{Lam}). They are
\ba\label{crit}
\fl
\gamma_{1, 2}=\frac12 \left(-dx_{c} \pm \sqrt{d^2x_c^2
+4m_{21}} \right)\,,~~~
m_{21}=-\frac{1}{2c_{2}} \left[4(d-3)c_1x_{c}^2+(d-1)\right]\,.
\ea
If $(d-3)c_{1}$ is positive, we find $m_{21}<0$ for $c_2>0$, 
which means that both $\gamma_{1}$ and $\gamma_{2}$ are
negative. Then this case corresponds to a stable attractor.
Meanwhile the dS solution is unstable for $(d-3)c_{1}>0$
and $c_2<0$, since one has $\gamma_{1}>0$ and $\gamma_{2}<0$.

When $(d-3)c_1<0$ there exist two dS fixed points given by 
equation (\ref{two}). 
For the first critical point in equation~(\ref{two}), one has 
$m_{21}<0$ for $c_{2}>0$ and $m_{21}>0$ otherwise.
Therefore the dS solution is stable for $c_{2}>0$
and a saddle point otherwise.
Similarly we find that the second critical point in 
equation (\ref{two}) is stable for $c_{2}<0$ 
and a saddle point otherwise.

{}From the above argument it is clear that the stability of the dS
solutions crucially depends upon the signs of $c_{1}$ and $c_{2}$.


\section{Linear dilaton case}\label{linear}

{}From now on we shall study the 
cosmological evolution in the presence of a dynamically 
varying scalar field. 
For the dilatonic action in the string frame, 
the condition (\ref{deSidilaton}) for the existence 
of the dS solution yields
\be
\label{eq1}
\fl 
\lambda d(d-3)c_{1}x^4-4d\lambda c_{1} v x^3+d(d-1)x^2
-2dvx+v^2-3a_4\lambda v^4=0\,.
\ee
This suggests that dS solutions may exist when 
$v=\dot{\phi}$ is a non-zero constant. 

Using $\dot{v}=0$ in equation (\ref{veq}), one gets
\be
\label{eq2}
\fl
-\lambda d(d+1)c_{1}x^4-4d\lambda a_{4}v^3x-d(d+1)x^2
+2dvx-v^2+3a_4\lambda v^4=0\,.
\ee
{}From the above two equations we find that the dS 
solution satisfies
\ba
\label{eq3}
2\lambda c_1x^3+2\lambda (c_{1} vx^2+a_{4}v^3)+x=0\,.
\ea
We note that the Minkowski solution ($x=0$ and $v=0$)
also satisfies equations (\ref{eq1}) and (\ref{eq2}).

When $d=3$ equations (\ref{eq1}) and (\ref{eq3}) give
\ba
\label{eq4}
x^2=\frac{-v^2(9\lambda a_{4}v^2+1)}
{6(1+2\lambda c_1 v^2)}\,.
\ea
For the bosonic correction with a GB term one has 
$\lambda=1/4$, $a_{4}=-1$ and $c_{1}=2$.
Then by combining equations~(\ref{eq3}) and (\ref{eq4}), 
we obtain the following dS solution 
with a linearly changing dilaton:
\ba\label{val}
x_c=0.62\,,\qquad v_c=1.40.
\ea
This agrees with the result in Ref.~\cite{CHC}.
It is clear from equation (\ref{eq4}) that the dS solution does not 
exist for $a_{4}=0$ for the GB coupling ($c_1=2)$.
If we allow for a negative value of $c_{1}$, 
it is possible to have a dS solution even when $a_{4}=0$. 
Note that these results do not depend on $c_2$.

\begin{figure}
\bc
\includegraphics[width=3.2in]{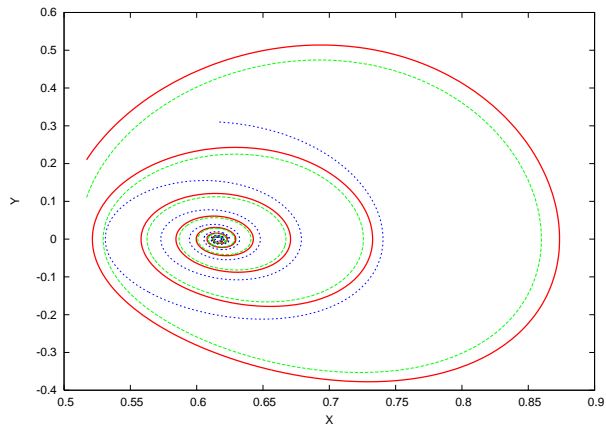}
\ec
\caption{\label{fig2}
The phase portrait (plot of $y$ versus $x$) of the system in the linear dilaton case with $d=3$,  
$c_{1}=2$, $\lambda=1/4$, $a_4=-1$ and $c_{2}=-1$.
Trajectories starting anywhere in the phase space converge at  $(x_c, y_c, z_c)=
(0.62, 0, 0)$ for a constant value of $v =1.40$. The fixed point corresponds to a stable de Sitter solution.
}
\end{figure}

The issue of the stability of such solutions
is presented in more details in the Appendix
(including the case of $c_2=0$). 
A rather counter-intuitive result is the following.
Although the dS solution is insensitive to the actual value 
of $c_2$, its stability depends upon the sign of $c_2$. 
More precisely, the solution is stable against 
linear perturbations for $a_4<0$, $c_1>0$, $c_2\geq 0$, 
whereas for negative $c_2$ one of the eigenvalues has a positive 
real part and the solution is unstable. 
We have checked other situations for $d=3$, which can be
summarized as follows 
(we considered only real solutions with $x_c>0$):
\begin{itemize}
\item $c_1>0$, $a_4\geq 0$: There is no real solution.
\item $c_1>0$, $a_4<0$: Real solutions ($a_4<-(9\lambda v^2)^{-1}$) are stable only for $c_2\geq 0$.
\item $c_1<0$, $a_4> 0$: Real solutions ($c_1<-(2\lambda v^2)^{-1}$) are stable only for $c_2\geq 0$.
\item $c_1<0$, $a_4 \leq 0$: Real solutions are stable only for $c_2 \leq 0$. See figure \ref{fig2}.
\end{itemize}

We should stress that the dS solutions we found appear in the presence of 
the $\xi$-dependent terms.
In the low-energy regime where the effect of higher curvature terms is neglected, 
we obtain a logarithmic evolution of the dilaton corresponding to 
the pre/post-big-bang solution in pre-big-bang cosmology \cite{GV,GV02}.
Later we shall briefly discuss the case of  
a logarithmic dilaton in the string frame.


\subsection{Impact of the barotropic fluid}

In the case of vanishing coupling $Q$, the above stability analysis holds 
as long as we assume a standard equation of state characterized by $w>-1$. 
In the presence of a phantom fluid (diverging $z_c$) the eigenvalue 
$\gamma_2=-d(1+w)x_c$ is always positive for $x_c>0$ and 
all solutions are unstable (see the Appendix). 
If $Q \neq 0$ the dS solutions are stable when
\be
Q \leq d(1+w)\frac{x_c}{v_c},
\ee
where we assumed $v_c>0$.
Therefore the phantom fluid with positive $Q$ 
is always unstable.


\subsection{Cosmological constant $(w = -1)$}

When $w=-1$ and $z=\Lambda=\,$const, the left-hand side of 
equation (\ref{eq1}) has an extra term $-2\Lambda e^u$. 
This term vanishes or diverges depending on the sign of $u$. 
In order to find a non-trivial dS solution with cosmological constant, 
one has to assume that the dilaton field $u$ approaches 0 asymptotically, 
$v\to$0, $u\to 0$ as $t \to \infty$. 
In this regime, equation (\ref{eq2}) gives
\be \label{lamv}
x^2_c=\frac{\Lambda}{d}=-\frac{1}{\lambda c_1},
\ee
which is positive for $c_1<0$. 
The value of the cosmological constant is fixed by the choice of the coefficients.
We ran our numerical code for various values of $c_1< 0$ and $c_2$ 
and checked that the dS solution in four dimensions 
is unstable against linear perturbations.
Note that this behaviour does not coincide with what has been 
found for a fixed dilaton in section \ref{w-1}. 
This is because the dynamical equations are affected by the presence 
of the scalar field even before reaching the asymptotical regime.

A quick inspection of the dynamical equations shows that in the 
Einstein frame there are no dS solutions, except under the same assumption 
adopted for the cosmological constant case ($u \to 0$ asymptotically).


\section{Logarithmic scalar field and the late time cosmological evolution}\label{modu}

In the first part of this section we shall consider the cosmological dynamics in the 
presence of the compactification modulus under the assumption that 
the dilaton is fixed.
This means that the analysis is identical both in the Einstein
and string frame. In section \ref{logdil} the logarithmic dilaton in the string frame
will be studied as well.
In order to keep our discussion as general as possible, 
we shall not fix the dimension $d$ of the target space 
except at the end, where the four-dimensional case $d=3$ 
will be inspected among the others. 
The equations of motion for the modulus action corresponding 
to Eqs.~(\ref{modu1})-(\ref{modu2}) read
\ba
\fl 
\dot{x} = y,\label{modeq1}\\
\fl 
\dot{y} = \frac{2z+3v^2/2-d(d-1)x^2-8dc_1\dot{\xi} x^3-2d(d-3)c_1\xi x^4
-2d c_2\xi y(2\Xi x+2dx^2-y)}{4dc_2\xi x},\nonumber\\\label{modeq2}\\
\fl 
\dot{v} = \frac{2d}{3}\xi'[(d+1)c_1 x^4+4 c_1 x^2 y+c_2 y^2]-dxv-\frac{2Q}{3}z,\label{modeq3}\\
\fl 
\dot{z} = \left[-dx(1+w)+Qv \right]z.
\label{modeq4}
\ea
While only derivatives of $\xi$ appear in the equations of motion 
for $d=3$ and $c_2=0$ (GB case), there are non-vanishing 
contributions of $\xi$ itself for general coefficients $c_i$.
When $c_2=0$, the equations of motion for $x$ and $v$ read, 
from equations (\ref{frc20}) and (\ref{veq}),
\ba
\dot{x} &=& \frac{\dot{z}+3v\dot{v}/2-dc_1x^3[4\ddot{\xi}+(d-3)\dot{\xi}x]}
{dx [(d-1)+12c_1\dot{\xi}x+4(d-3)c_1\xi x^2]},
\label{eqc201}\\
\frac{\dot{v}}{v^2} &=& \frac{2d}{3}c_1\xi'\frac{x^2}{v^2}
[(d+1)x^2+4\dot{x}]-d\frac{x}{v}-\frac{2Q}{3}\frac{z}{v^2},
\label{eqc202}
\ea
while the Friedmann equation is
\be\label{femo}
d(d-1)\frac{x^2}{v^2}-\frac32-2\frac{z}{v^2}+2dc_1\xi \frac{x^3}{v^2}[4\Xi+(d-3)x]=0.
\ee
When finding solutions one can use the equations for $c_2\neq 0$ 
and then set $c_2=0$ at the end, although equations (\ref{eqc201})--(\ref{femo}) 
are required for numerical purposes. 
In addition $c_1$ can be set equal to 1 and absorbed in the definition of $\xi_0$, 
so that the coefficient $c_2$ is the only free parameter of the higher-order Lagrangian.

Following the approach of \cite{ART,KSS}, 
we search for future asymptotic solution of the form
\begin{numparts}\ba
x &\sim& \omega_1 t^\beta,\qquad y \sim \beta\omega_1 t^{\beta-1},\label{asso1}\\
u &\sim&  u_0+\omega_2\ln t,\qquad v \sim \frac{\omega_2}{t},\label{asso2}\\
\xi &\sim& \frac12 \xi_0e^{u_0}t^{\omega_2},\label{asso3}\\
z &\sim& z_0t^{Q\omega_2}\exp\left[-\frac{d(1+w)\omega_1}
{\beta+1}t^{\beta+1}\right],\qquad \beta\neq -1,\label{zne1}\\
z &\sim& z_0t^{\alpha},\qquad \beta= -1,\label{asso5}
\ea\end{numparts}
where the barotropic index $w$ is constant and
\be\label{Qw}
\alpha \equiv Q\omega_2-d(1+w)\omega_1.
\ee
We define 
\be
\tilde{\delta}\equiv \frac12 c_1\xi_0e^{u_0},
\ee
so that in any claim involving the sign of $\tilde{\delta}$ ($\delta$) a positive $c_1$ 
coefficient is understood. In order to find a solution in the limit $t \to +\infty$, 
one has to match the exponents of $t$ to get algebraic equations 
in the parameters $\beta, \omega_i,c_2$ and $\tilde{\delta}$. 

We shall consider the following four cases: 
\begin{enumerate}
\item A low-curvature regime in which $\xi$ terms are subdominant at late times.
\item An intermediate regime where some terms in the equations of motion, either coupled to $\xi$ or not,
are damped. 
\item A high-curvature regime in which $\xi$ terms dominate.
\item A solution of the form (\ref{asso1})--(\ref{asso5}) for the \emph{full} equations of 
motion.\footnote{Low- and high-curvature regimes may be viewed even as 
``weak coupling'' and ``strong coupling'' models, respectively.
These correspond to (late-time) stages at which higher-order terms 
become either subdominant or dominant because of the relative magnitude 
of the theory's coefficients, evaluated at some moment of the cosmological evolution.
Quotation marks are in order since actually the curvature is not related to the modulus 
expectation value $u$; however, in the Friedmann and modulus equations the higher-order 
term $\xi\rho_c$ dominates both when $\xi\gg 1$ and $\rho_c\gg 1$, and 
the two descriptions are equivalent (with $a_4=0$). 
The varying coupling picture may be confusing in our context, since curvature terms 
may dominate because of their asymptotical
time dependence, for natural values of their couplings, while the 
weak/strong coupling approximation is assumed to hold in 
a time interval $\Delta t$ where the dynamical variables 
have the above time dependence.} 
\end{enumerate}

Our results will be in agreement with \cite{ART} (and \cite{KSS}, 
taking into account our extra $\sqrt{2/3}$ 
factor in the normalization of $v$) for $d=3$ and $z=0$.

In order to keep contact with observations, we shall compare the 
obtained values of $\omega_1$ of the geometrical ($z=0$) modulus solutions 
with the estimate of the deceleration parameter 
from supernovae data \cite{rie04}, which does not depend on the Einstein equations governing the background evolution \cite{TR} (while the relation between the deceleration parameter and the dark energy barotropic index does). 
Solutions in agreement with observations might be used to describe 
the recent evolution of the universe. We shall not consider the logarithmic dilaton solution since it does not belong to the runaway dilaton models of \cite{DP1,DP2} and may be in conflict with gravity experiments, although we have not checked this explicitly.

As we will see later, one has $\beta=-1$ for all except 
one case.  Then $\omega_1$ is given
in terms of the deceleration parameter $q \equiv \epsilon-1$:
\be
\omega_1 =\frac{1}{1+q}.
\ee
The first thing to note is that equation (\ref{asso1}) 
can describe only cosmologies with $q>-1$, that is, 
the effective phantom regime is avoided.

At late times (redshift $z\sim 0$), the observational data 
give the constraint $-1.0 \lesssim q_0\lesssim -0.5$ \cite{rie04}, 
where $q_0=q(z)|_{z\sim 0}$. 
This results in the condition for $\omega_1$
\be
\omega_1^{(0)} \gtrsim 2,\\
\ee
if the obtained solution describes the present universe.

Of course this constraint is loosened if we use 
the $3\sigma$ observational bounds, 
but we need to exclude the region $q<-1$ for the above argument.
For $-1.3 \lesssim q_0\lesssim -0.2$ 
(after exclusion of the extremal negative region), 
one has $\omega_1^{(0)} \gtrsim 1.2$.

In order to allow a super-accelerating phase ($q<-1$), 
we can shift the origin of time so that $t\to t-t_s$. 
For positive increasing time $t<t_s$, the solution 
\be
x=\frac{\omega_1}{t-t_s},
\ee
is expanding when $\omega_1<0$. 
The constraints on the parameters can be found by considering 
the previous asymptotic behaviour for $t\to 0$, 
that is near the singularity at $t_s$. 
Then the previously discarded $3\sigma$ 
region $-1.3 \lesssim q_0\lesssim -1$ corresponds to 
\be\label{phan}
\omega_1^{(0)} \lesssim -57.8.
\ee
In order to satisfy this bound, however, the parameters of the 
solutions will require rather unnatural adjustments.

We shall not discuss the stability of these solutions 
since they will hold at times $t \sim t_s$, 
that is, only near the sudden future singularity.


\subsection{Structure of the solutions and behaviour of the perfect fluid}

Before proceeding, let us explain the general structure of the solutions. 
Each of the equations of motion, under the substitution (\ref{asso1})--(\ref{asso5}), 
is of the form
\be\label{geom}
\sum_i \left(g_i^+t^{\alpha_i^+}+ g_i^-t^{-\alpha_i^-}\right)=0,
\ee
where $g_i^\pm=g_i^\pm(\beta,\omega_1,\omega_2,c_1,c_2,\xi_0,\dots)$ 
are the functions of the parameters of the theory, and 
$\alpha_i^\pm=\alpha_i^\pm(\beta,\omega_1,\omega_2,Q,w)$ 
are non-negative exponents. 
In the limit $t \to \infty$, an asymptotic solution with 
\be\label{gsol}
\alpha_i^+=0,\qquad \alpha_i^->0,
\ee
is $\sum_i g_i^+=0$. If the $g_i^+$ coefficients do not depend on $\xi_0$, 
then the solution represents a low-curvature regime. 
If $\partial g_j^+/\partial \xi_0\neq 0$ for some $j$, 
then we will say that the solution describes an intermediate regime, 
when some (but not all the) higher-order terms survive. 
If $\partial g_i^+/\partial \xi_0\neq 0$ for all $i$, 
only the higher-order terms contribute and the solution 
is in a high-curvature regime. 

In the limit $t\to 0$, the solution (\ref{gsol}) becomes $\sum_i g_i^-=0$, 
with eventually (in fact, only in the intermediate regime) some 
subdominant $g_i^-t^{-\alpha_i^-}$ terms dying away. 
Therefore all the above cases are interpreted in the complementary way, 
and low (high) curvature solutions become high (low) curvature.

The solution satisfying $\alpha_i^+=\alpha_i^-=0$ is 
valid at all times and the only distinction between solutions 
with or without future singularity is the sign of $\omega_1$.

As regards the contribution of the barotropic fluid, 
when $\beta=-1$ we can parametrize its time dependence in equation (\ref{geom}) as
\be
g_z(t)=z_0t^{\alpha-\alpha_*},
\ee
where $\alpha$ is defined in equation (\ref{Qw}) and $\alpha_*$ is 
a constant determined by the kind of solution (low/high curvature, exact, etc.) 
one is searching for.

We can adopt two perspectives. 
In the first one, we do not fine-tune the fluid parameters and impose that $g_z(t)$ 
vanishes asymptotically (purely geometrical solution).
In this case, $z\to 0$ asymptotically in the equations of motion 
and $\alpha_*-\alpha \in \{\alpha_i^-\}$. 
{}From equation (\ref{gsol}) it follows that $\alpha<\alpha_*$. 
In the second one, the expansion is driven by the perfect fluid and $g_z=\,$const, 
that is, $\alpha-\alpha_*\in \{\alpha_i^+\}$, and $\alpha=\alpha_*$. 
We summarize these constraints in the damping/matching condition
\be\label{DMcon}
\alpha\leq\alpha_*.
\ee
When the inequality holds, the fluid decays away 
in the limit $t\to +\infty$ and is consistently discarded.

When equation (\ref{DMcon}) is an equality, the fluid contributes 
to the asymptotical equations together with other geometrical terms
and affects the dynamics of the system. 
However, this does not mean that the energy density $z(t)$ increases
in time, since $\alpha$ can be negative. 
In this case the barotropic fluid can be regarded 
as a model for dark energy with 
\ba
& & \epsilon = \frac{1}{\omega_1}=\frac{d(1+w_{\rm eff})}{2},\\
& & w_{\rm eff} \equiv \frac{2w-Q\omega_2}{2+Q\omega_2}.
\ea
Then constraints on $q$ would result in bounds for $w_{\rm eff}$. 
In the asymptotic limit $t\to +\infty$, solutions exist (and do not burst out) 
only for the effective barotropic index $w_{\rm eff}>-1$. However, 
in this section we shall constrain only the geometrical solutions 
with a scalar field $\phi$, 
postponing a more complete discussion on dark energy to the next section.

To keep the notation clear, we shall write explicitly the time dependence of fluid 
terms and discuss separately the damping and matching cases. 
In the last section we shall compare numerically evolved dark energy scenarios
to the analytical solutions below.
In some cases the fluid will be dominating or decaying and 
either the matching or the damping condition will hold at the infinite future, 
with fixed constant parameters $\omega_1$ and $\omega_2$. 
At intermediate times, 
where avoidance or incoming of a big rip become manifest, 
there is enough freedom in varying $w$ and $Q$ to span a wide range
of dark energy phenomena, since $d g_z/dt\neq 0$ in general.


\subsection{Low-curvature and intermediate regimes} \label{lcsol}

The solution in a low-curvature regime corresponds to the one 
where $\xi$-dependent terms are neglected.
The asymptotic future solution we consider is ($d\neq 1$)
\be \label{grsol}
\beta=-1,\qquad \omega_2<2,
\ee
together with the constraints
\ba
\omega_1 &=& \frac{1}{d}-\frac{2Qz_0t^{\alpha+2}}{3d\omega_2},\label{grsol1}\\
3\omega_2^2 &=& 2d(d-1)\omega_1^2-4z_0t^{\alpha+2},\label{grsol2}\\
\alpha &\leq& -2,\label{Qeq}
\ea
so that the energy density $z(t)$ decreases in time. $\xi$ terms decay away regardless of the value of $c_2$. 
According to equation (\ref{Qeq}), if $\omega_2>0$ and the barotropic fluid 
is phantom-like ($w<-1$), then $Q$ should be negative and of high enough magnitude
in an expanding universe 
($\omega_1>0$).\footnote{Constraints on the sign of $Q$ will depend on 
the sign of $\omega_2$. Assuming $v<0$ would simply flip sgn$(Q)$ in these considerations.} 
The reader can check other situations. 
In the damping case $\alpha<-2$, 
the solution is $\omega_1=1/d$ and $\omega_2=\pm\sqrt{2(d-1)/(3d)}$. 
This case (for $d=3$) was labelled $A_\infty$ in Ref.~\cite{ART}.

In the marginal case $\alpha=-2$, the scale factor expands for $Q/\omega_2<3/(2z_0)$.
This requires either an ordinary fluid with $Q\omega_2>-2$ or a phantom fluid with $Q\omega_2<-2$. 
When $Q=0$ one has $w=1$. In addition equation (\ref{grsol2}) states that $\omega_1^2>2z_0/[d (d-1)]$. 
We have numerically checked that this solution exists for suitable values of $Q$.

Another solution, valid in an intermediate regime and only for the $c_2=0$ and $\omega_1>0$ case, is
\be\label{minsol}
\fl
\beta=-2,\qquad \omega_2=5,\qquad Q\leq-2/5,\qquad \omega_1^3
=\frac{1}{16d\tilde{\delta}}\left(15-2Qz_0t^{5Q+2}\right),
\ee
for a non-vanishing fluid.
We require the condition $\tilde{\delta}>0$ in order to have an expanding scale factor 
$a(t)\sim a_0\exp(-\omega_1/t)$. 
Actually this goes from 0 to $a_0$, that is, it reaches Minkowski spacetime
asymptotically. If the fluid decays, then one recovers the $C_\infty$ solution 
of Ref.~\cite{ART} with $d=3$ and $z=0$. 
Interestingly, the coupling $Q$ cannot vanish when $z\neq 0$.

For both the low-curvature and intermediate regimes, there exist no 
solutions corresponding to (\ref{asso1})--(\ref{asso5}) 
in the cosmological constant case ($w=-1$), since the $\Lambda/v^2$ 
term diverges asymptotically.

\begin{figure}
\bc
\includegraphics[width=3.2in]{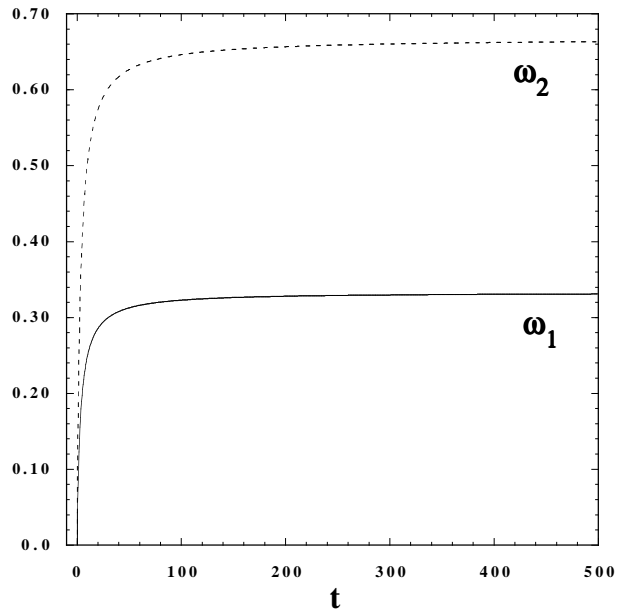}
\ec
\caption{\label{fig3}
Evolution of $\omega_{1}$ and $\omega_2$ for
$c_2=0$, $d=3$, $z=0$, and $\xi_{0}=2$ without 
a barotropic fluid.
We find that the system approaches the low-curvature solution (\ref{grsol})
characterized by constant $\omega_{1}$ and $\omega_{2}$ 
($\omega_{1}=1/3$ and $\omega_{2}=2/3$).
}
\end{figure}

We performed numerical simulations for $c_{2}=0$, $z=0$, and 
$\omega_2>0$ with general initial conditions determined by the Hamiltonian constraint.
With either positive or negative values of $\tilde{\delta}$, 
we found that in the asymptotic future 
the solutions tend to approach the general-relativistic (GR) solution (\ref{grsol}) 
rather than the intermediate solution (\ref{minsol}). 
One example is shown 
in figure~\ref{fig3}, corresponding to the case without 
barotropic fluid and $d=3$.
The asymptotic solution is actually described by 
constant values of $\omega_{1}$ and $\omega_{2}$ characterized by 
equations (\ref{grsol1}) and  (\ref{grsol2}), i.e., $\omega_{1}=1/3$ 
and $\omega_{2}=2/3$. 
The $\delta<0$ case is consistent with the result in Ref.~\cite{ART}, 
where non-singular cosmological 
solutions were constructed by joining the future GR solution (\ref{grsol}) 
with the past intermediate solution (\ref{minsol}).
In fact we numerically found that the asymptotic past solutions are well 
described by equation (\ref{minsol}) for negative $\delta$.
One can either choose a positive or negative sign of $\dot{\phi}$
in order to obtain non-singular cosmological solutions.
When $\delta>0$, we found that the past solutions are not 
described by equation (\ref{minsol}), 
which is associated with the fact that singularity-free solutions
do not exist in this case \cite{ART}.

It is easy to check the stability of low-curvature solutions without fluid 
analytically. Using equations (\ref{eqc201})--(\ref{femo}) for $z=0$, 
we obtain, after dropping $\xi$-dependent terms,
\be
\dot{x}\approx -dx^2,
\ee
which gives
\be
\dot{\omega}_1=\frac{\omega_1(1-d\omega_1)}{t}.
\ee
Considering perturbations $\delta\omega_1(t)$
around the critical point $\omega_1^{(0)}=1/d$, we obtain
\be
\delta \omega_1 \sim 
\frac{1}{t^{2d \omega_1^{(0)}-1}} = \frac{1}{t}.
\ee
Since $v$ and $x$ are proportional to each other, the perturbations
in $\omega_2$ decay as $\delta \omega_1$, thereby showing the stability
of the GR solution. 
In the presence of a fluid with $Q=0$, it can be shown that the solution 
is stable again, provided that $w>-1$.

The solution with a sudden future singularity at $t_s$ corresponds to 
\be
\beta=-1,\qquad \omega_2>2;
\ee
and an eventual cosmological constant term damps away. 
The parameter constraints are equations (\ref{grsol1}) and (\ref{grsol2}). 
In the marginal case
\be
\alpha = -2,
\ee
the scale factor expands for $Q>3\omega_2/(2z_0)>3/z_0$.
This requires a phantom fluid with $w+1=(2+Q\omega_2)/(d\omega_1)<0$. 
We have numerically checked that this solution exists for suitable values of $Q$.


\subsection{High curvature regime}

When higher-order terms become important we can consider the asymptotic future solution
\be\label{rem}
\beta=-1,\qquad \omega_2>2,\qquad \alpha\leq \omega_2-4,
\ee
together with constraint equations
\ba
&& \tilde{\delta}d\omega_1^2[(d+1)\omega_1^2-4\omega_1+c_2]
-Qz_0t^{\alpha-\omega_2+4}=0,\label{rem1}\\
&& d\tilde{\delta}\omega_1^2 [(3-d)\omega_1^2-4\omega_2\omega_1
+c_2(2d\omega_1+2\omega_2-3)]+z_0t^{\alpha-\omega_2+4}=0.\label{rem2}
\ea
In the GB case ($c_2=0$) the solution corresponding to a decaying 
fluid ($\alpha<\omega_{2}-4$) is 
\be
\omega_1 = \frac{4}{d+1},\qquad 
\omega_2 = \frac{3-d}{d+1},
\label{remsol}
\ee
which contradicts the condition (\ref{rem}) in any dimension. Therefore in the GB scenario 
with $\omega_2\neq 4$ only the marginal case $\alpha=\omega_2-4$ is allowed.

When
\be
\omega_{2}=4,\label{sollam}
\ee
the cosmological constant term ($w=-1$) survives; an additional fluid term would be trivial since $\alpha=0$.
The constraints of this solution are
\ba
&& (d+1)\omega_1^2-4\omega_1+c_2=0,\\
&& d\tilde{\delta}\omega_1^2 [(3-d)\omega_1^2
-16\omega_1+c_2(2d\omega_1+5)]+\Lambda=0.
\ea
In the GB case ($c_{2}=0$) we find
\ba
\omega_1 &=& \frac{4}{d+1},\label{sollam1}\\
\Lambda  &=& \tilde{\delta} d (5d+1) 
\left(\frac{4}{d+1}\right)^4. \label{sollam2}
\ea
The presence of a non-vanishing cosmological constant is mandatory 
for the consistency of this solution. 
For example one has $\Lambda= 48\tilde{\delta}$ for $d=3$. 
It is remarkable that equation (\ref{sollam}) is the only situation 
where a non-zero cosmological constant is not only allowed
but even required in the GB case. 

\begin{figure}
\bc
\includegraphics[width=3.2in]{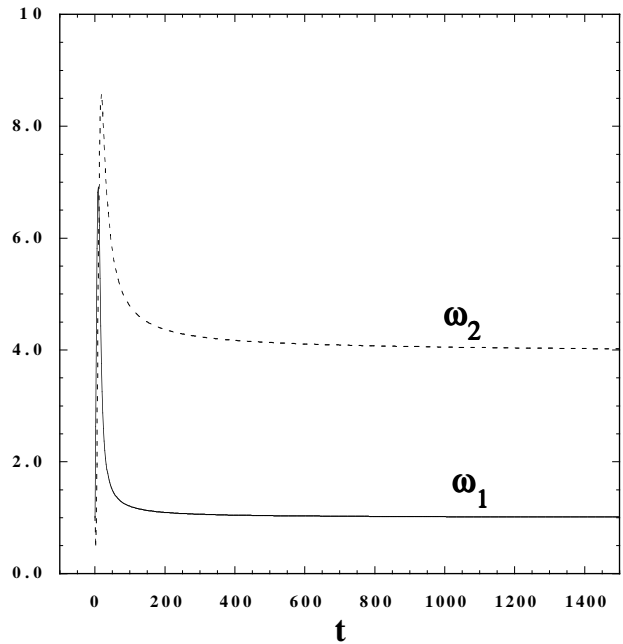}
\ec
\caption{\label{fig4}
Variation of $\omega_{1}$ and $\omega_2$ for
$c_2=0$, $d=3$, and $\xi_{0}=2$ in the presence of 
a cosmological constant given by $\Lambda= 48\tilde{\delta}$.
The asymptotic behavior of $H$ and $\dot{\phi}$
is characterized by $\beta=-1$, $\omega_{1}=1$ and 
$\omega_{2}=4$ as estimated analytically.}
\end{figure}

In order to check whether the solutions really approach the one given 
by  equation~(\ref{sollam}), we run our numerical code 
for $d=3$, $c_{2}=0$, and 
$\tilde{\delta}>0$ with $\Lambda=48\tilde{\delta}$.
As illustrated in figure~\ref{fig4}, the future asymptotic solutions are 
described by equations (\ref{sollam}) and (\ref{sollam1}) 
with constant $\omega_{1}$ (=1) and $\omega_{2}$ (=4). 
We chose several different initial conditions and found 
that the solution (\ref{sollam}) is a stable attractor.
We also considered the case of a negative cosmological constant, but 
the solution (\ref{sollam}) is found to be unstable.

When $c_2\neq 0$, not all values and signs of the coefficient will give 
real expanding solutions. Since
\be\label{usesol}
\omega_1^\pm=\frac{2\pm \sqrt{4-(d+1)c_2}}{d+1},
\ee
the reality condition reads $c_2\leq 4/(d+1)$.
If $c_2<0$, only the $\omega_1^+$ solution is expanding, whereas both solutions 
are for $0<c_2\leq 4/(d+1)$. 
Again, the sign of the cosmological constant will depend upon $\delta$.

Numerically, the solution of this system for $d=3$ approaches the analytical value of $\omega_1^+$ 
(and the associated $\omega_2$) for positive $\Lambda$. 
The $\omega_1^-$ branch is found to be unstable regardless of 
the sign of the cosmological constant. 
Therefore the only stable solution is $\omega_1^+$ with $\Lambda>0$.

There are non-trivial positive solutions even with $\Lambda=0$, 
corresponding to the critical case 
\be
\omega_1=\frac{2}{d+1},\qquad c_2=\frac{4}{d+1}.
\ee
We chose several different initial conditions and found no stable attractor. 
Since we did not check it for an extensive set of initial conditions, 
we cannot claim anything definite regarding the stability of this solution.

The sudden future solution corresponds to
\be \label{grsolsf}
\beta=-1,\qquad \omega_2<2,
\ee
which describes a high-curvature regime where the $\xi$-dependent terms
are dominant (a cosmological constant term is suppressed). 
The constraints on the parameters are 
equations (\ref{rem1}), (\ref{rem2}) and
\be
\alpha\geq \omega_2-4,
\ee
where $z\to 0$ when the last equation is an inequality. 
In this case the solution (\ref{remsol}) is contracting for $c_2=0$.  
Conversely, for $z=0$ and $c_2\neq 0$ the only expanding solution 
is $\omega_1^-$ given in equation (\ref{usesol}) for $c_2<0$, together with 
\be
\omega_2=3-d\omega_1^-.
\ee
%


\subsection{Exact solution}

An exact solution which is valid at all times is
\be\label{exsol}
\beta=-1,\qquad \omega_2=2,
\ee
together with the constraints on $\omega_1$:
\ba
\fl 
& & 2\tilde{\delta} d\omega_1^2[(d+1)\omega_1^2-4\omega_1+c_2]
-6d\omega_1+6-Qz_0t^{\alpha+2}=0,\label{exso1}\\
\fl 
& & 2\tilde{\delta} d\omega_1^2[(d-3)\omega_1^2+8\omega_1
-c_2(2d\omega_1+1)]+d(d-1)\omega_1^2-6-2z_0t^{\alpha+2}=0,\label{exso2}
\ea
and equation (\ref{Qeq}). 
The cosmological constant term is forbidden in this case.

Let us consider the case with a vanishing fluid ($z\to 0$). 
After some manipulations, the above equations can be written as
\ba
&& 4\tilde{\delta}\omega_1^2[(d-1)\omega_1+2-dc_2]+(d-1)\omega_1-6=0,\label{useful}\\
&& 2\tilde{\delta}\omega_1[d(d+1)\omega_1^2-2(d+1)\omega_1+4-dc_2]-(5d+1)=0.
\ea
When $c_2=0$, these equations completely fix the parameters of the theory. 
In four dimensions, for instance, the only real root of the above equations is 
\be\label{solex}
(\omega_1,\tilde{\delta})=(0.21,13.53),
\ee
in agreement with Ref.~\cite{ART} (solution $B_\infty$). 
Note that the original coupling $\delta$ is not fine-tuned, although $\tilde{\delta}$ is, 
since we have a freedom in choosing $u_0$. 
There is no solution with a sudden future singularity.

Numerically we found that the exact solution (\ref{exsol}) is not stable.
As we already pointed out in section \ref{lcsol}, the future stable attractor 
for $\Lambda=0$ corresponds to the GR solution given by equation~(\ref{grsol}) for 
both positive and negative $\delta$. The asymptotic past solution 
is described by equation (\ref{minsol}), with $\delta<0$.

In the case $c_2\neq 0$, from equation (\ref{useful}) 
one can show that a solution of the system is given by ($d\neq 1$)
\be\label{parsol}
\omega_1=\frac{6}{d-1},\qquad c_2=\frac8d,\qquad \tilde{\delta}=\frac{(d-1)^3}{48(d+2)}.
\ee
We checked numerically that this solution is unstable.

Other (real) solutions, with either $\omega_1>0$ or $\omega_1<0$, 
can be found by fixing the parameters known \emph{a priori} from the theory, 
that is, $c_2$ and $\tilde{\delta}$.


\subsection{Logarithmic dilaton}\label{logdil}

With little effort, we can check whether equations (\ref{asso1})--(\ref{asso5}) 
are a solution of the dilatonic model in string frame for $t\to +\infty$. 
It is easy to find that the only possible case is $\beta=-1$, 
corresponding to a low-curvature regime 
where $\xi$-dependent terms are damped away. 
Then the model reduces to a scalar-tensor theory in Einstein gravity. 

In the presence of a barotropic fluid, 
the matching/damping condition reads $\alpha\leq -2-\omega_2$, i.e.,
\be
(1+Q)\omega_2-d(1+w)\omega_1+2\leq 0,
\ee
while the constraints on the parameters read, with the above matching condition,
\ba
d\omega_1^2-d\omega_1+\omega_2 +(1+Q)e^{u_0}
z_0t^{\alpha+\omega_2+2}=0,\label{logdi1}\\
d(d-1)\omega_1^2-2d\omega_1\omega_2+\omega_2^2 
-2e^{u_0}z_0t^{\alpha+\omega_2+2}=0.\label{logdi2}
\ea
When $\alpha= -2-\omega_2$, the solutions are
\ba
\fl 
 \omega_1^\pm = \pm \sqrt{\frac{1}{2d}\left[1-2(1+Q)e^{u_0}z_0+\sqrt{1+4(1-Q)e^{u_0}z_0}\right]}\,,\\
\fl 
 \omega_2^\pm = \frac12 \left\{\vphantom{\frac12}-1-\sqrt{1+4(1-Q)e^{u_0}z_0}\right.\nonumber\\
\lo \pm \left.\vphantom{\frac12}\sqrt{2d\left[1-2(1+Q)e^{u_0}z_0+\sqrt{1+4(1-Q)e^{u_0}z_0}\right]}\right\},\\
\fl 
\tilde{\omega}_1^\pm = \pm \sqrt{\frac{1}{2d}\left[1-2(1+Q)e^{u_0}z_0-\sqrt{1+4(1-Q)e^{u_0}z_0}\right]}\,,\\
\fl 
\tilde{\omega}_2^\pm = \frac12 \left\{\vphantom{\frac12}-1+\sqrt{1+4(1-Q)e^{u_0}z_0}\right.\nonumber\\
\lo \pm \left.\vphantom{\frac12}\sqrt{2d\left[1-2(1+Q)e^{u_0}z_0-\sqrt{1+4(1-Q)e^{u_0}z_0}\right]}\right\}.
\ea
Without fluid ($Q=0=z_0$), the expanding solution reads
\ba
\omega_1^+ &=& \frac{1}{\sqrt{d}},\\
\omega_2^+ &=& \sqrt{d}-1,
\ea
in agreement with Ref.~\cite{GV}. 
One has $\omega_1^+\approx 0.58$ and $\omega_2^+\approx 0.73$
for $d=3$. This solution is found to be stable numerically.

In the presence of a cosmological constant [$z_0t^\alpha=\Lambda$ and $Q=0$ 
in equations (\ref{logdi1}) and (\ref{logdi2})], we get 
$\omega_1=-2/(1+d)$ and $\omega_2=-2$, which describes a contracting 
universe.

The solution with a sudden future singularity at $t_s$ and 
$\beta=-1$ corresponds to a high-curvature regime where 
$\xi$-dependent terms are dominant.
In this case the constraints read
\ba
\fl 
3\lambda a_4 \omega_2^4-\lambda d\omega_1^2[(d-3)c_1\omega_1^2
-4c_1\omega_1\omega_2+c_2(2\omega_2-2d\omega_1+3)]
+2e^{u_0}z_0t^{\alpha+\omega_2+4}=0,\\
\fl 
2\lambda a_4 \omega_2^3(d\omega_1-\omega_2-3)+
\frac{\lambda}{2}\left\{d\omega_1^2[(d+1)c_1 \omega_1^2-4c_1\omega_1+c_2]
+a_4\omega_2^4\right\}+Qe^{u_0}z_0t^{\alpha+\omega_2+4}=0.\nonumber\\
\ea
When $\Lambda\neq z\neq 0$, the damping/matching condition 
for the fluid is
\be
(1+Q)\omega_2-d(1+w)\omega_1+4\geq 0.
\ee
When the fluid vanishes or decays, we verified that there exist non-trivial solutions 
for suitable choices of parameters, 
both without and with a cosmological constant
($z=\Lambda$ and $\omega_2=-4$ in the above equations). 
For instance, in four dimensions and in the GB case 
($d=3$, $c_1=2$, $c_2=0$, $a_4=-1$), 
the only expanding solution is $\omega_1=-2.87$ and $\Lambda =-188.88$. 
One can find other solutions by varying the parameters $c_1,c_2,a_4,\Lambda$.


\subsection{Constraints from the current universe}

We now compare the observational constraints on $\omega_1$ 
for the recent evolution of the universe with the modulus solutions found 
in the previous section ($d=3$).
We consider the situation in which a 
perfect fluid is vanishing asymptotically.
The results are summarized in Table \ref{tab1} 
with the 68\% CL bound. 
We discarded the intermediate regime or Minkowski solution.

\begin{table}[ht]
\begin{center}
\begin{tabular}{l||c|c}\br
Solutions  $t\to \infty$ & $\omega_1^{(0)}$ (68\% CL)     & Stability \\ \mr
Low curvature &                	& Yes       \\
High curvature, $c_2=0$    &           & $\Lambda>0$ \\ 
High curvature, $c_2\neq 0$, $\Lambda=0$    &              								   &          \\
High curvature, $c_2\neq 0$, $\Lambda\neq0$ & $\omega_1^+$, $c_2\lesssim -8$ & $\omega_1^+$, $\Lambda>0$\\ 
Exact, $c_2=0$                              &                                &\\
Exact, $c_2\neq0$, equation (\ref{parsol})  &     Yes                        & \\ \br
\end{tabular}
\caption{\label{tab1} 
Constraints on modulus solutions 
in the asymptotic future for the Hubble parameter $H=\omega_1 t^{-1}$, 
vanishing fluid, and $d=3$. 
Blank entries are excluded by experiments or numerical analysis.}
\end{center}
\end{table}

We find that the logarithmic modulus solution with ghost-free GB parametrization and no extra fluid does not provide 
a viable cosmological evolution in the current late-time universe. However, we will see in the next section
that the GB case in the presence of a cosmological fluid may display interesting features as regards the future evolution of the universe.
At the 99\% confidence level, the low redshift constraint on $c_2$ 
for the high curvature, 
$\Lambda\neq 0$ solution can be relaxed up to $c_2\lesssim -1$. 
Therefore we have shown that there are models which can in principle explain the present acceleration 
without using the dark energy fluid.

The situation becomes more complicated 
in the presence of a barotropic fluid.
The low-curvature solution can describe the very recent universe 
if $Q$ is negative and non-vanishing. The other cases crucially 
depend upon the interplay between all the theoretical parameters.

As regards the models with a sudden future singularity, the bound equation (\ref{phan}) 
generally requires severe fine tunings for the parameters of the solutions. 
For instance one needs the condition $Qz_0 \gtrsim 10^3$ for
the low-curvature solution with $\omega_2=4$ and $d=3$.
In the high-curvature case (\ref{grsolsf}) with a vanishing fluid, we 
require that $|c_2|\gtrsim 10^4$.
In the exact case ($c_2\neq 0$),
$|\tilde{\delta}|\lesssim 10^{-6}$. 
In the dilaton case with a cosmological constant, $|\Lambda| \gtrsim 10^7-10^9$.

However, we found a remarkable theoretical result. 
This kind of models with $c_2=0$ (Gauss--Bonnet case) 
always requires a non-vanishing contribution from the barotropic fluid for consistency. 
In the low-curvature case, even a positive coupling $Q$ is necessary. 
In the dilaton case, a non-vanishing cosmological constant must be present
if $z=0$.


\section{Gauss--Bonnet modulus gravity and the dark energy universe}\label{dark}

In the previous section we compared cosmological solutions obtained 
in second-order gravity with the present evolution of the universe. 
In this section we shall adopt another perspective, namely, 
we will investigate the phenomenology of the full equations of motion 
as models of future evolution of the dark energy universe. At the infinite future, 
the properties of such models will agree with some of the above solutions. 

In the typical phantom scenario ($w<-1$), the 
energy density of the fluid 
continues to grow and the Hubble rate eventually exhibits 
a divergence at finite time (big rip).

It is expected that the effect of higher-order curvature corrections
becomes important when the energy density grows up to the 
Planck scale. In fact, it was shown in \cite{NOT,NO1,NO2} that 
inclusion of quantum curvature corrections coming from 
conformal anomaly can moderate the future singularities.

Here we would like to consider the effect of $\Or(\alpha')$ string
quantum corrections when the curvature of the universe increases
in the presence of a phantom fluid. We shall concentrate on the modulus case 
with $\xi$ given by equations (\ref{ximodu1}) and (\ref{ximodu2}).
Our main interest is the cosmological evolution in four 
dimensions ($d=3$) in the presence of a GB term ($c_1=2$, $c_2=0$) with 
$a_{4}=0$. We assume that the dilaton is stabilized, so that there are no long-range forces to take into account except gravity.

{}From the discussion in section \ref{modu}, we found that the growth of 
the barotropic fluid is weaker than that of the Hubble rate 
when equation (\ref{Qeq}) is satisfied as an inequality.
This condition is not achieved for a phantom fluid 
when the coupling $Q$ between the fluid and the field $\phi$ 
is absent ($Q=0$).

\begin{figure}
\bc
\includegraphics[width=3.2in]{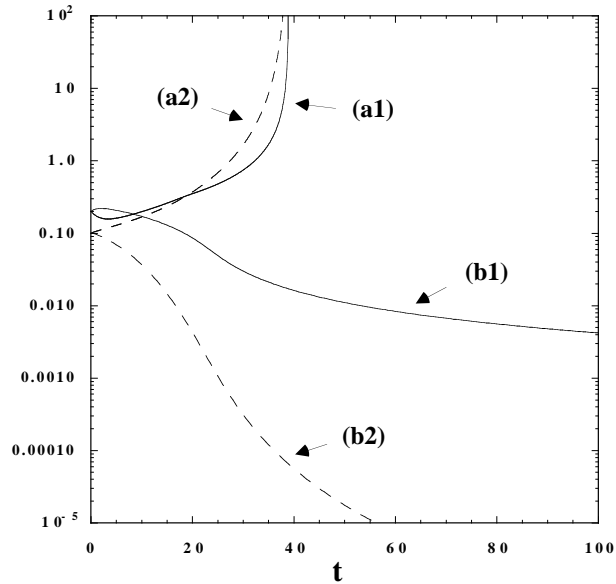}
\ec
\caption{\label{fig5}
Evolution of $H$ and $\rho$ with $\xi_{0}=-2$, $w=-1.1$ for (a) 
$Q=0$ and (b) $Q=-5$. We choose initial conditions as $H_{i}=0.2$, 
$\phi_{i}=2.0$ and $\rho_{i}=0.1$.
The curves (a1) and (b1) represent the evolution of $H$ for
$Q=0$ and $Q=-5$, respectively, while the curves (a2) and (b2)
show the evolution of $\rho$ for corresponding $Q$.
We find that the big rip singularity is avoided when the negative 
coupling $Q$ is present.
}
\end{figure}

We ran our numerical code by varying initial conditions of 
$H$, $\phi$ and $\rho$ in the full equations of motion.
When $\delta<0$, we numerically found that 
the solutions approach a big rip singularity for $Q=0$
and $w<-1$ (see figure~\ref{fig5}).
Meanwhile the condition (\ref{Qeq}) can be satisfied for negative $Q$
provided that $\omega_{2}$ is positive.
In figure~\ref{fig5} we show the evolution of $H$ and $\rho$
for $Q=-5$ and $w=-1.1$. In this case $\rho$ decreases 
faster than $t^{-2}$, which means that 
the energy density of the fluid eventually becomes negligible 
relative to that of the modulus. Therefore the universe approaches the low-curvature
solution given by equation~(\ref{grsol}) at late times, thereby 
showing the avoidance of big rip singularity even for $w<-1$.
By substituting the asymptotic values $\omega_{1}=1/3$ and 
$\omega_{2}=2/3$ in equation~(\ref{Qeq}), the condition for 
decaying fluid reads $Q<3(w-1)/2=-3.15$.
We checked that the big rip singularity 
can be avoided in a wide range of the parameter space  
for negative $Q$. These results do 
not change even for smaller values of $\delta$ such as 
$|\delta|=\Or(1)$
(corresponding to $|\xi_0| = 0.1$).

When $Q$ is positive, the condition (\ref{Qeq}) is not fulfilled 
for $\omega_{2}>0$. However our numerical calculations show that 
$\dot{\phi}$ becomes negative even if $\dot{\phi}>0$ initially.
We found that the system approaches the low-energy regime  
characterised by $\omega_{1}=1/3$ and $\omega_{2}=-2/3$.
Since $\omega_{2}<0$, the big rip
singularity can be avoided even for positive $Q$.
In fact we numerically checked that the Hubble rate 
continues to decrease as long as the condition (\ref{Qeq})
is satisfied in the asymptotic regime.

\begin{figure}
\bc
\includegraphics[width=3.2in]{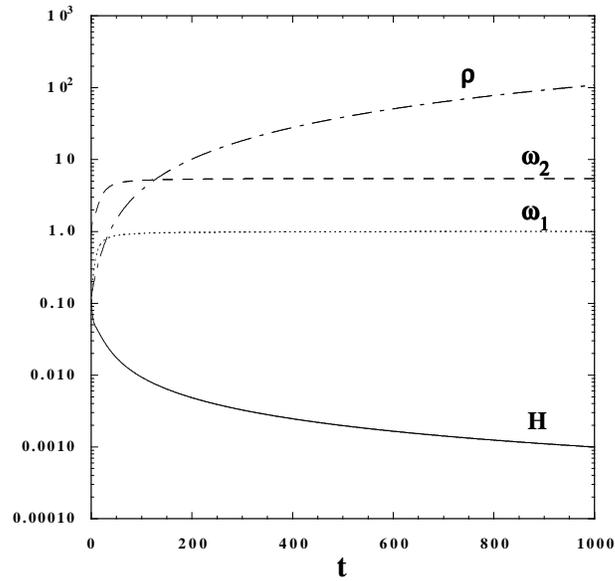}
\ec
\caption{\label{fig6}
Variation of $H$, $\rho$, $\omega_{1}$ and $\omega_{2}$ with 
$\xi_{0}=+2$, $w=-1.5$ and 
$Q=0$. Initial conditions are chosen to be $H_{i}=0.2$, 
$\phi_{i}=2.0$ and $\rho_{i}=0.1$.
The Hubble rate continues to decrease while the energy density 
$\rho$ increases.
}
\end{figure}

When $\delta>0$, there is another interesting situation in 
which the Hubble rate decreases in spite of the increase of the energy 
density of the fluid. This corresponds to the solution in the 
high-curvature regime in which the growing energy density $\rho$
can balance with the GB term ($\rho \approx 24H^3\dot{\xi}$ in 
the Friedmann equation). One example is illustrated in 
figure~\ref{fig6}; the big rip does not appear 
even when $w<-1$ and $Q=0$. Thus the GB corrections coupled with a scalar field $\phi$
provides us several interesting possibilities to get rid of this 
singularity.

\begin{figure}
\bc
\includegraphics[width=3.2in]{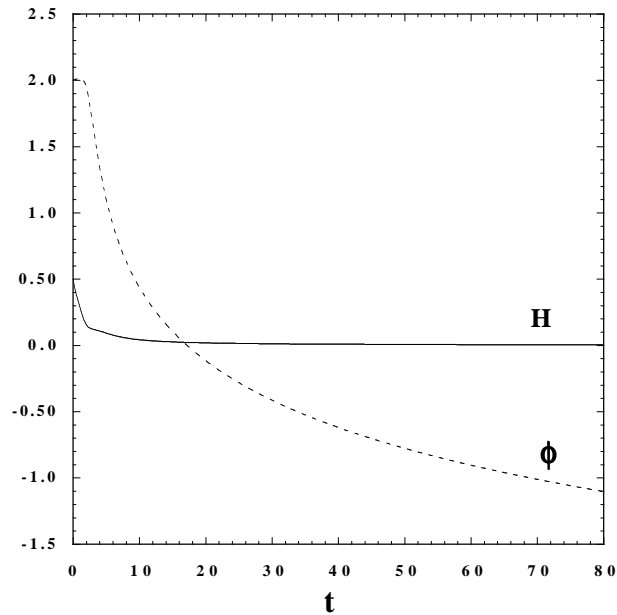}
\ec
\caption{\label{fig7}
Evolution of $H$ and $\phi$ with 
$\xi_{0}=-2$, $w=-0.8$ and 
$Q=+5$. Initial conditions are chosen to be $H_{i}=0.5$, 
$\phi_{i}=2.0$ and $\rho_{i}=0.3$.
We find that both $H$ and $\phi$ continue to decrease
even in the presence of positive $Q$.
}
\end{figure}

By examining the condition (\ref{DMcon}),  
one may think that the big rip singularity may appear even for 
$w>-1$ in the presence of the coupling $Q$.
However we found that this is not the case.
We show one example in figure~\ref{fig7},
corresponding to $w=-0.8$ and $Q=+5$.
It turns out that $\dot{\phi}$ becomes negative
even when $\dot{\phi}>0$  initially, as seen in the figure.
We also checked that this property holds even for negative 
$Q$. This suggests that the system tends to evolve so that the 
condition 
\be
\alpha<0\qquad {\rm when}\qquad Q\neq 0,
\ee
is satisfied. Then the presence of the coupling $Q$ does not 
lead to the big rip singularity for a non-phantom
fluid ($w>-1$).

\section{Conclusions} \label{concl}

In this paper we have studied cosmological solutions in the presence 
of second-order curvature corrections 
and their application to the dark energy universe.
Our starting action (\ref{act}) is based upon the low-energy 
effective string theory with a barotropic perfect fluid
coupled to a scalar field $\phi$.
This is regarded as either the dilaton field 
characterising the strength of the string coupling, or 
a modulus field corresponding to the compactification radius of a dimensionally reduced theory.
We adopted a general form of the second-order correction terms, 
given by equation (\ref{Lc}), which includes the case of 
the Gauss--Bonnet curvature invariant.
We have clarified the property of cosmological solutions in three 
classes of models: ($i$) fixed scalar field (section \ref{dS}),
($ii$) dilaton field (sections \ref{linear} and \ref{modu}), and 
($iii$) modulus field (section \ref{modu}).

When the field $\phi$ is fixed (model ($i$)), the existence of pure geometrical 
de Sitter solutions requires a spatial dimension $d\neq 3$,
and that the parametrization of the second-order curvature correction is different from 
the GB one.
Therefore these solutions are not realistic when applied to inflation 
or dark energy. Nevertheless we studied the classical stability of such 
solutions in the presence of a perfect fluid or a cosmological 
constant for completeness.

If the dilaton field is not fixed (model ($ii$)), it is possible to construct pure
dS solutions in the string frame provided that the evolution 
of the dilaton is linear in time ($\dot{\phi}={\rm const}$).
In the case of $d=3$ spatial dimensions we obtained 
the dS solution given by Eq.~(\ref{val}) for the GB case, 
which was found to be stable.
We also clarified the stability of other cases different from the 
GB parametrization.
Moreover we showed that the existence of a phantom fluid generally 
makes the dS solution unstable.

When a modulus field is present with fixed dilaton (model ($iii$)), 
there exist cosmological solutions in which the field exhibits 
a logarithmic evolution, 
see equations (\ref{asso1})--(\ref{asso5}).
We showed the existence of low-curvature solutions (\ref{grsol})
and Minkowski solutions (\ref{minsol}), which can 
be joined each other if the coupling constant $\delta$ given 
in equation (\ref{modu2}) is negative.
In addition we obtained an exact solution (\ref{exsol}) for the 
modulus system, but this is found to be unstable.
In the asymptotic future the solutions tend to approach the 
low-curvature one given by equation (\ref{grsol}) rather than 
the others, irrespective of the sign of the modulus-to-curvature coupling $\delta$.
We also constructed high-curvature asymptotic solutions (\ref{rem})
in the presence of a growing perfect fluid or a cosmological constant.
Finally, we placed constraints on the viability of modulus-driven 
solutions using the current observational data.
It is interesting that the GB parametrisation is excluded 
in any of the above mentioned regimes when a barotropic fluid 
is vanishing; see table \ref{tab1}.

In section \ref{dark} we solved the full equations of motion to simulate 
a dark energy universe in the future. The numerical results were then compared to
the solutions obtained in previous sections.
We used the modulus coupling with 
the GB curvature correction in 3 spatial dimensions.
In the presence of the coupling $Q$ between the field $\phi$
and the phantom fluid ($w<-1$), it is possible to consider 
situations in which the energy density of the fluid decays 
in the future. In fact we have numerically found that 
the big rip singularity can be avoided for the coupling $Q$
which satisfies the condition $Q\omega_{2}-3(1+w)\omega_{1}<-2$
asymptotically (see figure \ref{fig5}). This is actually achieved 
irrespective of the sign of $Q$ and the asymptotic solutions are 
described by the low-curvature one given by equation (\ref{grsol}).
When $\delta$ is positive, we also found that even for $Q=0$
the Hubble parameter can decrease inspite of the growth of 
the energy density of the phantom fluid.
Therefore the big rip can be avoided in this case as well, 
as illustrated in figure \ref{fig6}.
When the equation of state of the fluid is non-phantom ($w>-1$), 
we found that the presence of the coupling $Q$ \emph{cannot} give rise
to the big rip singularity, contrary to what one might expect when 
considering the dark energy continuity equation (see figure \ref{fig7}).

Another topic which is worth studying further is the behaviour 
of the asymptotic solutions at high redshift. 
For instance, when the deceleration parameter is parametrized 
as $q(z)=q_0+q_1z$,\footnote{A change in the parametrization 
can affect the bounds on $q$ at large redshift; see \cite{VTTET} and references therein.} 
supernovae observations give constraints for both $q_0$ (redshift $z\sim 0$) and $q_1$ ($z\sim 0.5$). 
Here we limited the discussion to the low-redshift evolution, 
since the experimental uncertainties on high-redshift data, together with 
the correlated data analyses, are more delicate to deal with.
This topic is still under investigation.

In the presence of a well-motivated underlying theory, one should specify the physical scale at which high-curvature terms with a non-GB parametrization arise and make sense, either at early times (high-energy inflation) or late times (dark energy around a big rip). This energy scale is related to the string coupling governing the perturbative higher-order corrections to the Einstein--Hilbert action. Although our main interest for the dark energy model of section \ref{dark} was only the Gauss--Bonnet case, we are still missing the precise theoretical setup in which a non-GB parametrization arise at high energies. This problem is still open and potentially rich of consequences.

Also, in order to get close to real-world cosmology, we implemented a matter 
contribution into the Lagrangian and interpreted it as a 
dark energy source. Although the dilatonic/modulus corrections are motivated by 
string theory (in particular with the GB parametrization), the origin 
of the matter term still lacks in theoretical background. 
A completely coherent picture of dark energy should clarify its origin from first principles. 
Nonetheless our results do not rely on a particular choice for the kind of fluid, 
except for the assumption of the continuity equation (\ref{drho}) and $w=\,$const. 
Field/string theory should also determine the coupling $Q$ between the scalar 
field and matter. In our examples of big rip avoidance, the coupling $Q$ is of order unity and
has ``natural'' values anyway (in the sense specified by Dirac).

We hope that this kind of phenomenology would be able to originate 
from a more precise formulation of the underlying physics.


\ack
The work of G C is supported by a JSPS fellowship.
S T thanks JSPS for financial support (No.\,30318802). M S thanks
M Zahid and Q N Usmani for their hospitality at Jamia physics department
during the period of his leave from Jamia Millia Islamia, New Delhi.


\appendix

\section*{Appendix: Stability of \lowercase{d}S solutions with varying dilaton}\label{stab}
\setcounter{section}{1}

The equations of motion (\ref{eoms1})--(\ref{zeq}), 
linearized under a linear perturbation of the vector 
$X\equiv (x,y,u,v,z)^t$, read $\delta\dot{X}=M\delta X$, 
where the $5 \times 5$ matrix elements $m_{i\!j}$ are
\ba  
\fl 
& &
m_{11}= m_{13} = m_{14} = m_{15}= 0,\qquad m_{12}=1, \\
\fl
& &
m_{21} = -\frac{d-1}{2\lambda c_2} + \frac{4c_1 vx}{c_2}
-dy-\frac{3(d-3)c_1 x^2}{2c_2}-\frac{ze^{u}}{d\lambda c_2x^2}
-\frac{3a_4v^4}{2dc_2x^2}-\frac{y^2}{2x^2}+\frac{v^2}{2d\lambda c_2x^2},\nonumber\\
\fl 
& &
m_{22} =v-dx+\frac{y}{x},\qquad m_{23}=z m_{25}=\frac{ze^{u}}{d\lambda c_2x},\\
\fl
& & 
m_{24}=\frac{1}{\lambda c_2}+\frac{6a_4 v^3}{dc_2x}+\frac{2c_1x^2}{c_2}-\frac{v}{d\lambda c_2 x}+y,\\
\fl
& & 
m_{31}= m_{32} = m_{33} = m_{35}= 0,\qquad m_{34}=1, \\
\fl
& & 
m_{41}=\frac{d(d+1)x-dv+2d(d+1)\lambda c_1x^3+4d\lambda 
c_1 xy+2d\lambda a_4v^3}{1-6\lambda a_4 v^2},\\
\fl
& & 
m_{42}=\frac{d(1+2\lambda c_1x^2+\lambda c_2 y)}{1-6\lambda a_4 v^2},
\qquad m_{43}=z m_{45}=\frac{Qze^{u}}{1-6\lambda a_4 v^2},\\
\fl
& & 
m_{44}= \frac{1}{1-6\lambda a_4 v^2}\left\{v-dx-6\lambda a_4 v^3+6d\lambda a_4 v^2x
-\frac{6\lambda a_4 v}{1-6\lambda a_4 v^2}\left[2dxv\vphantom{Qe^u}\right.\right.\nonumber\\
\fl
& &
\qquad -2dy-v^2-d(d+1)\lambda c_1x^4-d(d+1)x^2-(4c_1x^2+c_2y)d\lambda y\nonumber\\
\fl
& &
\qquad \left.\vphantom{\frac{A}{B}}\left.-4d\lambda a_4 v^3x+3\lambda a_4v^4-Qze^u\right]\right\},\\
\fl
& & 
m_{51}= -d(1+w)z,\qquad m_{52}=m_{53}=0,\\
\fl
& & 
m_{54}=Qz,\qquad m_{55}=-d(1+w)x+Qv,
\ea
where we have dropped the subscript $c$ for the background quantities. 
For the dS solution with $Q=0$, 
two eigenvalues of the matrix $M$ are given by $\gamma_1=0$ and $\gamma_2=m_{55}$. 
The other three can be extracted from
\be\label{eig}
\fl
\gamma^3-(m_{22}+m_{44})\gamma^2+(m_{22}m_{44}-m_{24}m_{42}-m_{21})
\gamma+(m_{21}m_{44}-m_{24}m_{41})=0.
\ee
In the four-dimensional case ($d=3$), by inserting $\lambda=1/4$, $a_{4}=-1$, $c_{1}=2$, 
and the values found in equation (\ref{val}), 
we find that the real part of the eigenvalues reads
\be
{\rm Re}(\gamma_3)=1.96,\qquad {\rm Re}(\gamma_4)=-2.40,\qquad {\rm Re}(\gamma_5)=-0.45,
\ee
for $c_2=1$, while
\be
{\rm Re}(\gamma_3)=-0.45,\qquad {\rm Re}(\gamma_4)= {\rm Re}(\gamma_5)=-0.22,
\ee
for $c_2=-1$. 
We have checked that the sign of the eigenvalues does not change
for different absolute values of $c_2$.

The GB case ($c_2=0$) must be treated separately. 
Together with the background equations (\ref{ueq})--(\ref{zeq}), 
we consider the (differentiated) Friedmann equation
\ba
\dot{x}=\frac{B_2}{B_1}\dot{v}+(vz+\dot{z})\frac{e^u}{B_1},
\ea
where
\ba
B_1 &\equiv& d(d-1)x-dv+2d(d-3)\lambda c_1 x^3-6d\lambda c_1 v x^2,\\
B_2 &\equiv& dx-v+6\lambda a_4 v^3+2d\lambda c_1 x^3.
\ea
At the dS point the linear perturbations of the $B_i$ coefficients are 
multiplied by vanishing factors.
Defining the reduced vector $\tilde{X} \equiv (x,u,v,z)^t$, 
it can be shown that the eigenvalues of the $4\times 4$ perturbation matrix 
$\tilde{M}$ at the dS point are
\ba
\gamma_{1,3} &=& 0,\qquad \gamma_3=
\tilde{m}_{55}=m_{55},\\
\gamma_4 &=& \tilde{m}_{11}+\tilde{m}_{44},
\ea
where
\ba
\tilde{m}_{11} &=& \frac{B_2}{B_1} \tilde{m}_{41}= \frac{B_2m_{41}}{B_1-B_2m_{42}},\\
\tilde{m}_{14} &=& \frac{B_2}{B_1} \tilde{m}_{44} = \frac{B_2m_{44}}{B_1-B_2m_{42}}.
\ea
Note that $\tilde{m}_{11}\tilde{m}_{44}-\tilde{m}_{41}\tilde{m}_{14}=0$. 
If $d=3$ and $\lambda=1/4$, 
the eigenvalue $\gamma_4$ is $-0.45$ for $a_{4}=-1$ and $c_{1}=2$, 
and negative for other choices of parameters. 
Therefore the dS solution is always stable. 
Note that we may derive the same results by just multiplying 
the Jordan equation (\ref{eig}) times $c_2$ and then setting $c_2=0$.
Since its coefficients are of the form ${\rm (constant)}_1+{\rm (constant)}_2/c_2$, 
only the ${\rm (constant)}_2$ terms survive and the above equation becomes linear. 
We have checked this shortcut explicitly.

In the presence of a cosmological constant, $z=\Lambda=\,$const, 
there exist dS solutions only in the asymptotical limit 
$u,v\to 0$ as $t\to \infty$ as explained in the text.
The only difference with respect to the previous expressions for $c_2=0$ 
is that they are evaluated at $v=0$ and 
$\tilde{m}_{14} \to \tilde{m}_{14}+\Lambda/(B_1-B_2m_{42})$, 
$\tilde{m}_{44} \to \hat{m}_{44}=\tilde{m}_{44}+\Lambda m_{42}/(B_1-B_2m_{42})$. 
In this case the equation for the eigenvalue 
has a non-vanishing constant coefficient and the solutions are
\ba
\gamma_{3,4} &=& \frac12 \left.\left[\left(\tilde{m}_{11}+\hat{m}_{44}\right)\pm 
\sqrt{\left(\tilde{m}_{11}+\hat{m}_{44}\right)^2+4m_\Lambda}\right]\right|_{v=0},\\
m_\Lambda &\equiv& \frac{\Lambda m_{41}}{B_1-B_2m_{42}}.
\ea
For negative $c_1$, one of the eigenvalues is always positive and the solution is unstable.


\end{document}